\documentclass[reprint,showpacs,showkeys,amsmath,amssymb,superscriptaddress,aps,prl]{revtex4-2}

\usepackage{graphicx}
\usepackage{float}
\usepackage{epstopdf}
\usepackage{bm}
\usepackage{amssymb}
\usepackage{amsmath}
\usepackage{bbold}
\usepackage[colorlinks=true, linkcolor=blue]{hyperref}
\usepackage{ulem}
\usepackage{color, soul}
\usepackage{xcolor}
\usepackage{multirow}
\usepackage[utf8]{inputenc}
\usepackage{xcolor, soul}
\sethlcolor{yellow}
\DeclareUnicodeCharacter{2212}{-}

\begin{document}
\preprint{Ojha {\it et al.}, Tailoring the Topological Hall Effect in Pt/Co/X (X = Ta, Re) thin films}

\title{Tailoring the Topological Hall Effect in Pt/Co/X (X = Ta, Re) thin films}

\author {Brindaban Ojha}
\author {Shaktiranjan Mohanty} 
\author {Bhuvneshwari Sharma}
\author{Subhankar Bedanta}
\email{sbedanta@niser.ac.in}
\address {Laboratory for Nanomagnetism and Magnetic Materials (LNMM), School of Physical Sciences, National Institute of Science Education and Research (NISER),  An OCC of Homi Bhabha National Institute (HBNI), Jatni 752050, Odisha, India}


\begin{abstract}
Electron transport combined with magnetism has gained more attention to the spintronics community in the last few decades. Among them, the topological Hall effect (THE), which arises due to the emergent magnetic field of a non-trivial object, is found to be a promising tool for probing the presence of skyrmions. A sizeable interfacial Dzyaloshinskii–Moriya interaction (iDMI) with reduced effective anisotropy can stabilize skyrmions in thin films. Recently, a large iDMI has been predicted in Pt/Co/Re thin film. Here, we investigate the influence of various magnetic interactions on the THE in perpendicularly magnetized Pt/Co/X (X = Ta, Re) thin films. The presence of skyrmions is confirmed via THE and magnetic force microscope (MFM) imaging. Notably, two distinct types of THE signals are observed in the different samples, which are explained using micromagnetic simulations. Our results reveal that exchange interaction, iDMI, effective anisotropy, and saturation magnetization contribute significantly in determining the variations in topological Hall resistivity behavior, which arise from different skyrmionic phases. These findings contribute to the development of novel material systems featuring different skyrmionic phases with potential applications in spintronics.
  
\end{abstract}


\maketitle
\section{Introduction}

\begin{figure*}
	\centering
	\includegraphics[width=0.7\linewidth]{"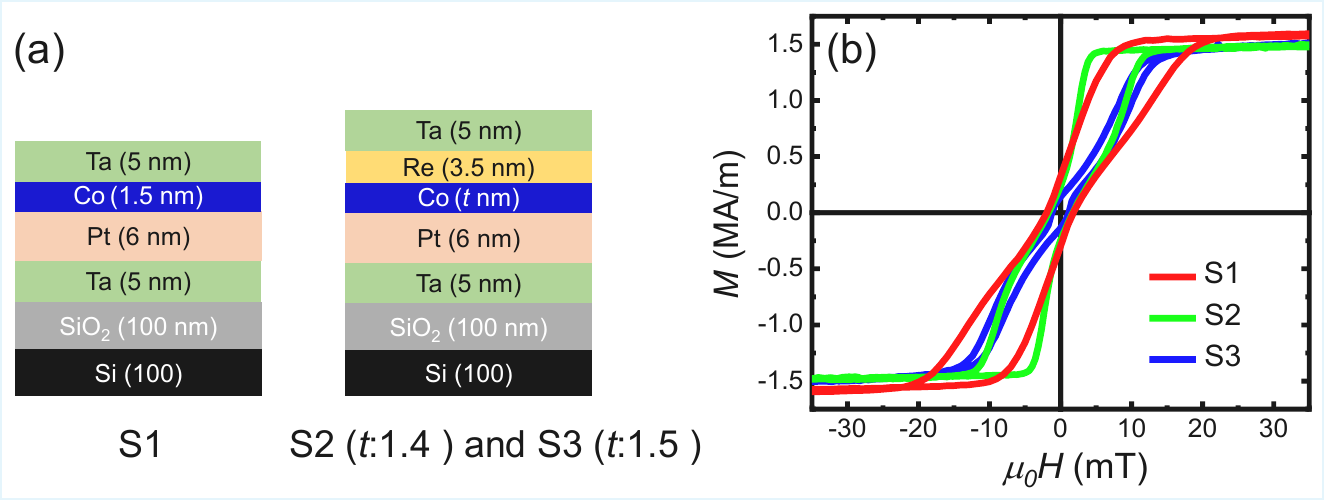"}
	\caption{\label{fig: wide}(a) Schematic of the sample structures. (b) OOP SQUID hysteresis loops of the samples S1-S3.}
	\label{fig:Fig_1}
\end{figure*}

Magnetic skyrmions are non-collinear, swirling, chiral topological spin textures. In recent decades, skyrmions have garnered significant interest in spintronics for their distinct topological properties, high thermal stability, nanoscale dimensions, and low critical current density for motion \cite{Fert2017,Karin2018}. It has been predicted that skyrmions are very promising candidates for enhancing next-generation data storage density, racetrack memory, logic devices, transistor-like devices, neuromorphic computing, etc. \cite{Fert2013,Karin2018,Luo2021,Tomasello2014}. Skyrmions stabilize in a thin film through a complicated interplay of exchange interaction, magnetic anisotropy, interfacial Dzyaloshinskii–Moriya interaction (iDMI), dipolar energy, and Zeeman energy \cite{Fert2013,Jiang2015,Luchaire2016,Hrabec2017,Ojha2023,bogdanov1994thermodynamically,bogdanov2020physical}. However, in specific circumstances, frustrated interaction, four-spin interaction, exchange bias, etc. also play a significant role in the formation of a skyrmion \cite{Romming2015,Heinze2011}. To meet the requirement of practical applications, skyrmions have been investigated in several ferromagnetic systems, ferromagnetic alloys, ferrimagnets, synthetic antiferromagnets, and recently in 2D materials \cite{Luchaire2016,Woo2018,Woo2016,Legrand2020,mallick2024driving,Wu2020}. The interaction of skyrmions with conduction electrons leads to exotic electrodynamics \cite{Schulz2012}. When a conduction electron moves through a slowly varying exchange field of chiral spin textures, i.e., skyrmions, chiral domain walls, etc., the spin of the electron follows spatially varying local magnetization in the adiabatic limit (or strong-coupling limit).  From the electron frame of reference, the electron experiences a time-dependent magnetic field due to a non-zero scalar spin chirality $\left(\chi_{i j k} \sim \boldsymbol{S}_i \cdot \boldsymbol{S}_j \times \boldsymbol{S}_k \neq 0\right)$, i.e., the solid angle occupied by the non-coplanar magnetization directions which leads to a Berry phase in real space \cite{Kanazawa2011,Huang2012}. Now, the Berry phase deflects the conduction electron perpendicular to the current flow direction and adds an additional contribution to the Hall signal, which is known as the topological Hall effect (THE). THE is a very useful and efficient tool to confirm the existence of chiral texture in a system. First, it was observed in non-centrosymmetric crystalline structures (B20-type bulk materials) \cite{Neubauer2009}. Later, it has also been explored in thin films having iDMI. Heavy metal (HM$_1$)/Ferromagnet (FM)/HM$_2$ is an ideal thin film system for stabilizing skyrmion and studying THE \cite{Soumyanarayanan2017,He2018,Mourkas2021,mohanty2024observation}. By choosing a proper combination of HM$_1$ and HM$_2$ across a FM, an additive iDMI can be achieved in a system that gives an additional degree of freedom in stabilizing the skyrmions. Recently, using first-principle calculations, it has been predicted that Pt/Co/Re system possesses a huge iDMI (iDMI constant, $D \sim$ 9.17 mJ/m$^2$) than any other thin film system like Pt/Co/W, etc. \cite{Jadaun12020,fakhredine2024huge}. Further, the experiment proved that Pt and Re induce opposite chirality, and the iDMI arising from Pt/Co and Co/Re interface added up \cite{Nomura2022}. The iDMI plays a pivotal role in stabilizing the skyrmion, which induces the THE in a system.  Furthermore, fine-tuning of the domain wall (DW) energy in a system is also essential to host the spin textures. A large positive energy collapses the skyrmions, whereas a large negative energy causes to destabilization of the colinear order (requires a higher magnetic field to achieve an isolated skyrmion).  The most efficient way of achieving this is by reducing the effective anisotropy energy of a system near to zero \cite{Hrabec2017,Mallick2022,Ojha2023}. Therefore, iDMI and anisotropy of a system are very crucial.  Previously, the effect of iDMI and anisotropy has been explored in skyrmion stabilization \cite{Behera2018}.  Further, the exchange and dipolar interactions also play an important role in determining the lowest energy of a system \cite{Thiaville2012,Rohart2013}. The influence of dipolar and exchange interactions on skyrmion stabilization is also investigated \cite{paul2020role,jefremovas2024role,vidal2017stability}. However, the influence of all these magnetic interactions on THE is less explored. The THE signal provides insight into the different topological phases present in the sample during magnetization reversal.

In this context, we have considered two model systems of Pt/Co/Ta and Pt/Co/Re thin films. It is well known that the Pt/Co interface induces a sizeable iDMI \cite{Woo2016,Belmeguenai2015}. However, very negligible iDMI (with the same sign as the Co/Pt interface) emerged from the top Co/Ta interface \cite{Shahbazi2019,Kashid2014}. So, the resulting iDMI in Pt/Co/Ta is expected to be smaller than a single Pt/Co interface \cite{Shahbazi2019,Kashid2014}. On the contrary, Co/Re exhibits a sizeable iDMI with the opposite sign of the Co/Pt interface. Therefore, in the case of the Pt/Co/Re system, both the interfaces contribute to induce an additive iDMI \cite{fakhredine2024huge,Jadaun12020,Nomura2022}. Thus, a higher iDMI is expected from the Pt/Co/Re system in comparison to Pt/Co/Ta. As both systems are expected to have different iDMI as well as other energies, the impact of these factors on the THE signal is studied in detail.  The topological Hall resistivity indicates the presence of skyrmions in the samples, as confirmed by magnetic force microscopy (MFM) imaging.  Further, the field-dependent topological Hall resistivity exhibits distinct behaviors in the two types of samples. A single peak or dip is found in the case of the Pt/Co/Ta system, whereas in Pt/Co/Re, the resistivity shows a simultaneous peak and dip (or double peak with opposite signs) during the field sweep.  To gain deeper insight into these Hall signatures, we have performed micromagnetic simulations. It is observed that exchange interaction, iDMI, effective anisotropy ($K_{eff}$), and saturation magnetization ($M_s$)  are crucial in defining topological Hall behaviour in a system.

\begin{figure*}
	\centering
	\includegraphics[width=0.7\linewidth]{"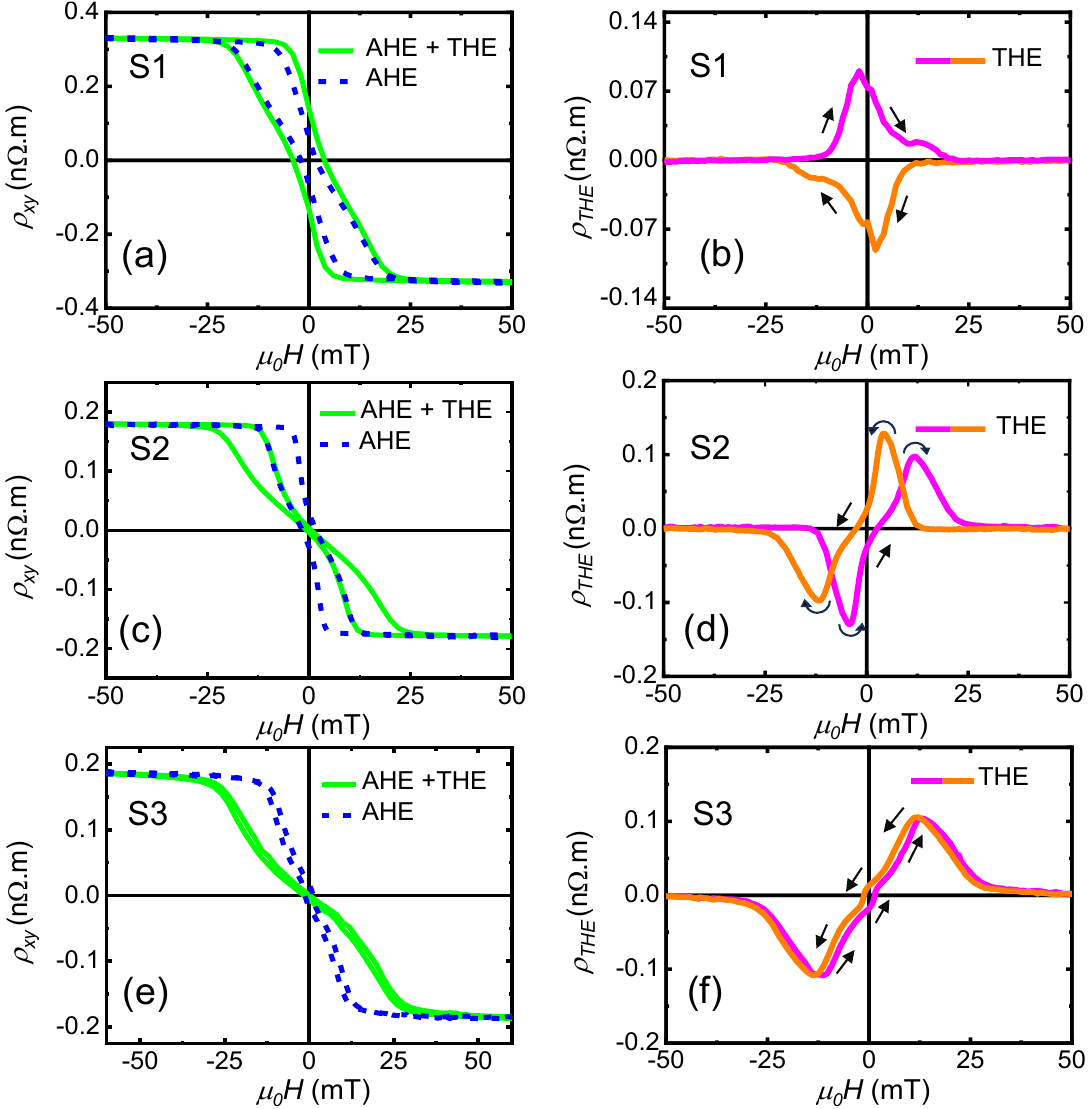"}
	\caption{(a), (c) and (e) AHE+THE and AHE contribution in $\rho_{xy}$ as a function of field for sample S1, S2 and S3, respectively. (b), (d) and (f) Plots of $\rho_{THE}$ vs field for sample S1, S2, and S3 respectively. The black arrow signs represent the field sweep direction.}
	\label{fig:Fig_2}
\end{figure*}

\section{Experimental details}

Three samples Ta(5)/Pt(6)/Co(1.5)/Ta(5), Ta(5)/Pt(6)/Co(1.4)/Re(3.5)/Ta(5) and Ta(5)/Pt(6)/Co(1.5)/Re(3.5)/Ta(5) have been prepared on 100 nm thermally oxidized Si substrates and named as S1, S2, and S3, respectively. The numbers in the parenthesis indicate the thicknesses of respective layers in nm. A schematic representation of these sample structures is shown in Fig. \ref{fig:Fig_1}(a). DC magnetron sputtering has been used to deposit the Ta, Pt, and Co layers, whereas RF magnetron sputtering has been used to prepare the Re layer with a base pressure $\sim 8\times10^{-8}$ mbar. Here, the Ta layer is used to facilitate the growth of Pt layer in (111) orientation, which induces out-of-plane magnetization in a thin Pt/Co system \cite{mallick2018relaxation}. Grazing incidence X-ray diffraction (GIXRD) measurements revealed a Pt/Co (111) reflection at 2$\theta$ $\approx$ 38$^{\circ}$ (see Fig. S1 in the supplementary information), indicating that Pt/Co films preferentially grow with a (111) texture \cite{kanak2007influence,Carcia1988,mangin2006current,parakkat2016tailoring,yildirim2022tuning}. The top Ta layer is deposited to prevent the thin films from oxidation. The structural parameters (e.g., thickness and roughness) of the multilayered samples are quantified via X-ray reflectivity (XRR) measurements (see Fig. S2 in the supplementary information). The substrate table has been rotated at 10 rpm during the deposition to achieve uniform layers. All the depositions are carried out at room temperature. Magneto optic Kerr effect (MOKE) based microscopy (manufactured by Evico-Magnetics, Germany) has been used to measure the hysteresis behaviour of the samples. The saturation magnetization ($M_s$) and effective anisotropy ($K_{eff}$) have been quantified via a superconducting quantum interference device (SQUID) magnetometer (manufactured by Quantum Design, USA). Skyrmion imaging has been performed via MFM (manufactured by Attocube, Germany) using commercially purchased magnetic tips (Nanosensor, SSS-MFMR). The topological Hall Effect measurements have been performed via a physical property measurement system (PPMS) (manufactured by Quantum Design, USA) in the $\pm$3 T field range in the Van der Pauw geometry. A small AC current density ($\sim$10$^6$ A/m$^2$) has been used for the transport measurements so that the spin textures do not perturb.  Micromagnetic simulations have been performed using MUMAX3 software \cite{Vansteenkiste2014}. The simulation area was 2 µm $\times$ 2 µm, with a cell size of  2 $\times$ 2 $\times$ 1.4 nm$^3$. The values of $M_s$ and $K_{eff}$ are varied within a range close to experimentally obtained values. The exchange constant, $A_{ex}$, is tuned in between 10--25 pJ/m, while the iDMI constant, $D$, ranges between 0.5--5.0 mJ/m$^2$ in the simulations. To ensure a realistic response, the simulation area is partitioned into 10 nm grains using random Voronoi tessellation. Periodic boundary conditions (PBC) are also applied in the simulations.

\section{Results and discussion}

\begin{figure*}
	\centering
	\includegraphics[width=0.8\linewidth]{"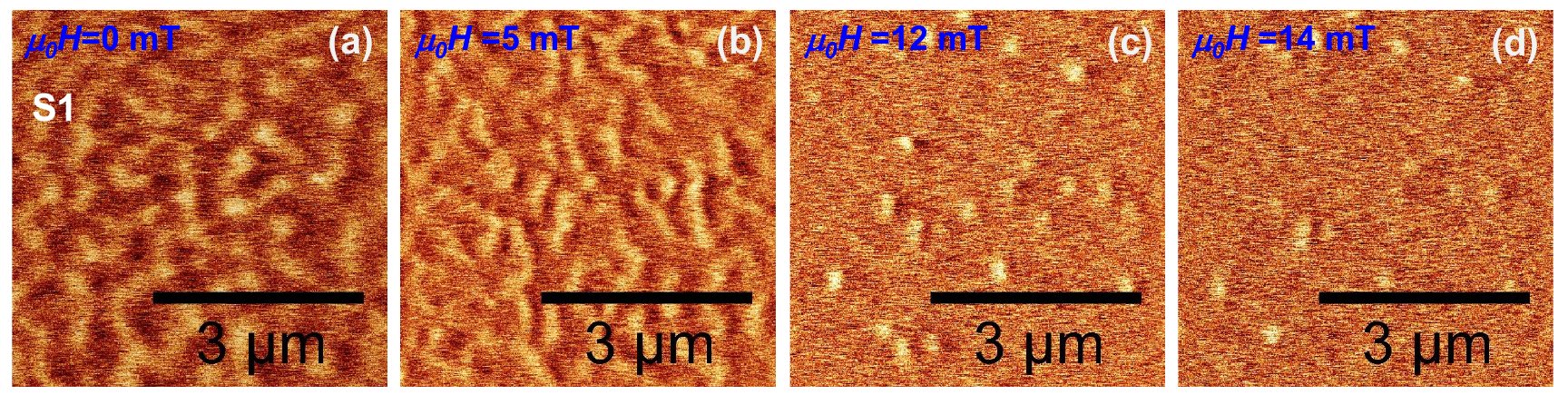"}
	\caption{(a-d) Field-dependent MFM images exhibiting the process of skyrmion formation. The isolated skyrmions are observed at $\mu_0 H$ $\ge$ 12 mT.}
	\label{fig:Fig_3}
\end{figure*}

In the presence of finite iDMI, the total energy density of an out-of-plane (OOP) system can be expressed as \cite{Thiaville2012,Rohart2013}:

\begin{widetext}
\begin{equation}
 E_{total} = E_{ex} + E_{dmi} + E_{ani} + E_{demag} + E_{zeem} 
 = A_{ex} (\nabla \vec{m})^2 + D \left[ m_z \cdot \nabla \vec{m} - (\vec{m} \cdot \nabla) m_z \right] + K_u(\vec{m} \cdot \hat{z})^2 - \frac{1}{2} \mu_0 M_s \vec{m} \cdot \vec{H_d} - \mu_0 M_s \vec{m} \cdot \vec{H}
 \label{eq:energy}
\end{equation}
\end{widetext}

In this expression,  $A_{ex}$  represents the exchange constant, $D$ is the iDMI constant, $K_u$ is the uniaxial anisotropy constant,   $\vec{m}$ reflects the magnetization vector, $m_z$ is the component of magnetization along $z$-axis,  $ M_s $ denotes the saturation magnetization, $\vec{H_d}$ defines the demagnetizing field, and $\vec{H}$ is the applied field.  The demagnetization energy, associated with shape anisotropy ($-\frac{1}{2} \mu_0 {M_s}^2$), modifies the effective anisotropy constant, as, $K_{eff}=K_u  -\frac{1}{2} \mu_0 {M_s}^2$ \cite{Mallick2022,Rohart2013}. Furthermore, the domain wall (DW) formation energy is given by \cite{Rohart2013,Pandey2023}:

\begin{equation}
 \sigma=4\sqrt{A_{ex} K_{\text{eff}}} - \pi D
 \label{eq:DWenergy}
\end{equation}
In non-collinear magnetic textures, at least one interaction, such as exchange coupling or uniaxial anisotropy, favors ferromagnetic (FM) alignment. While other competing interactions, such as Dzyaloshinskii-Moriya interaction (DMI) or dipolar interactions, promote non-collinear alignment of neighbouring spins \cite{Karin2018,Nagaosa2013}. Therefore, precise control of domain wall energy is crucial for stabilizing complex spin textures.

Fig. \ref{fig:Fig_1}(b) shows the OOP SQUID hysteresis loops, confirming that the easy axis of all the samples aligns in the OOP direction. The hysteresis loops of sample S3 exhibit a more slanted shape compared to S2, despite the relatively small variation in Co thickness between the two samples. In thin films, surface or interface anisotropy, and shape anisotropy are critical factors in determining the preferred magnetization direction. While surface or interface anisotropy favors perpendicular magnetization, shape anisotropy promotes in-plane magnetization \cite{Johnson1996}. Near the spin reorientation transition (SRT), the competition between perpendicular and in-plane anisotropies leads to an energy landscape where the two contributions are nearly degenerate. Consequently, the system becomes highly sensitive to minor perturbations, including variations in film thickness, temperature, defects, and external magnetic fields. Notably, the perpendicular anisotropy, primarily arising from interfacial effects, scales inversely with the thickness of the ferromagnetic layer (1/$t_{FM}$) \cite{Johnson1996}. As a result, even small changes in thickness can significantly modify the heights of the energy barrier, leading to distinct magnetization reversal mechanisms and pronounced variations in coercivity, remanence, and the overall shape of the hysteresis loop\cite{Ojha2023,Mallick2022}. Also, the presence of a low-remanence, slanted hysteresis loop indicates the existence of chiral spin textures \cite{Berges2022}, with the Co layer thickness tuned near to SRT in all samples. Now, in an OOP-magnetized thin film, increasing the FM thickness toward the SRT significantly reduces the effective anisotropy ($K_{eff}$), leading to a progressive slanting of the hysteresis loops \cite{Woo2016, Ojha2023,Mallick2022}. In this regime, strong iDMI further lowers domain-wall energy, destabilizing the uniform magnetization state for $D > \frac{4\sqrt{A_{ex} K_{\text{eff}}}}{\pi}$ and inducing chirality \cite{Rohart2013,Mallick2022}, as shown in eq. \ref{eq:DWenergy}. The values of $M_s$ and $K_{eff}$  of all samples are given in Table \ref{table:table1}. The $K_{eff}$ values have been determined using the relation, $K_{eff} = \frac{1}{2}\mu_0 H_{s}M_{s}$ where $\mu_0 H_s$ is the in-plane saturation field (as shown in Fig. S4 of the supplementary information).

\begin{figure}[h!]
	\centering
	\includegraphics[width=0.9\linewidth]{"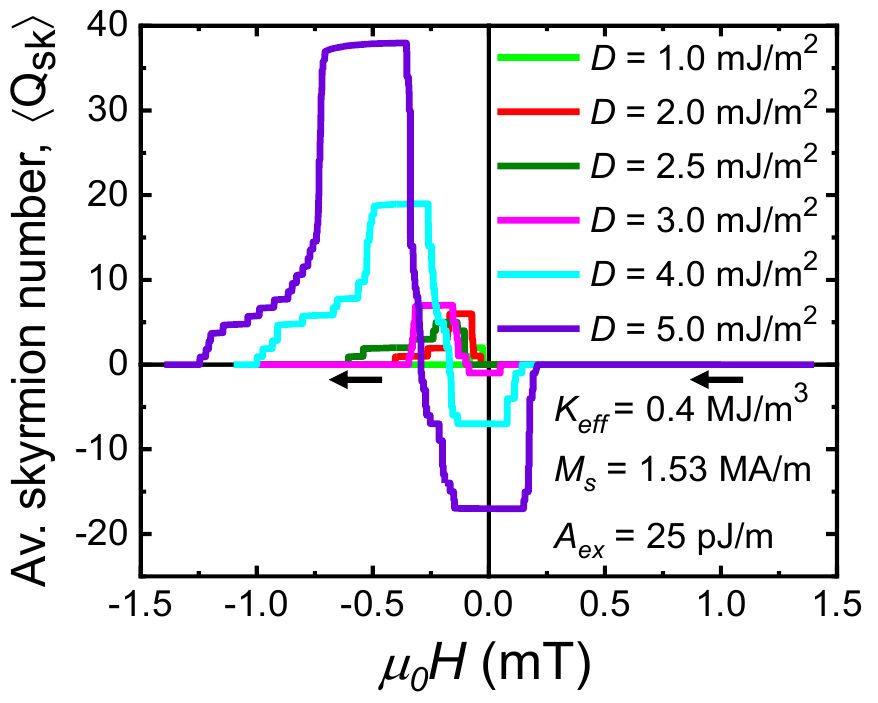"}
	\caption{ Plot of $\left\langle Q_{sk} \right\rangle$ vs $\mu_0 H$ at a constant $K_{eff}$, $M_s$, and $A_{ex}$ obtained from simulations. The black arrows represent the field sweep direction.}
	\label{fig:Fig_4}
\end{figure}

\begin{figure*}
	\centering
	\includegraphics[width=0.9\linewidth]{"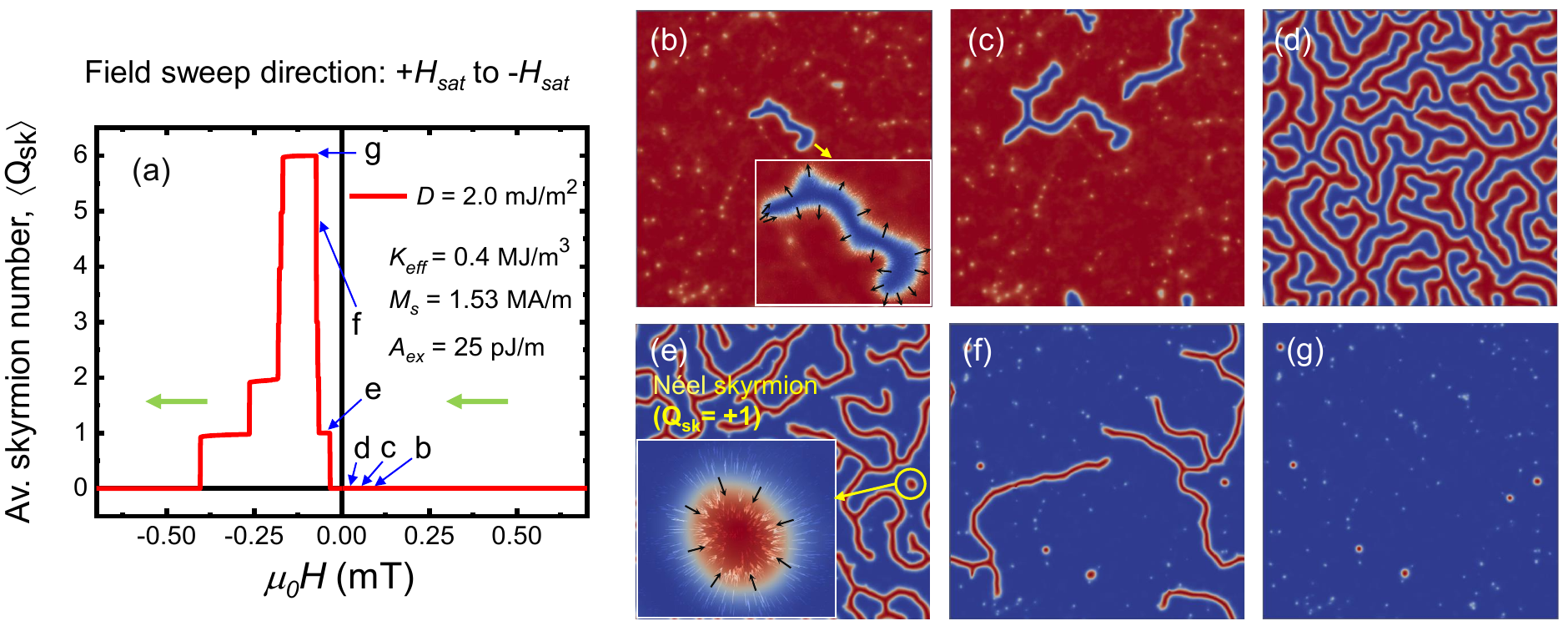"}
	\caption{(a) A single peak in $\left\langle Q_{sk} \right\rangle$ for $D$ = 2.0 mJ/m$^2$ (with $K_{eff}$ = 0.4 MJ/m$^3$, $M_s$=1.53 MA/m, and $A_{ex}$=25 pJ/m) obtained from simulations. The blue arrows refer to the corresponding domain structures as shown in (b-g). The green arrows represent the field sweep direction.}
	\label{fig:Fig_5}
\end{figure*}

To confirm the existence of the chiral textures in the samples, THE measurements have been performed. In a FM material, the total Hall resistivity ($\rho_{xy}$) is the sum of three contributions i.e., ordinary ($\rho_{OHE}$), anomalous ($\rho_{AHE}$), and topological ($\rho_{THE}$) Hall resistivities. The ordinary Hall effect (OHE) is linearly proportional to the applied field ($H$). The anomalous Hall effect (AHE) arises due to the time-reversal symmetry breaking in the FM and is linearly proportional to the magnetization. THE originated in a thin film due to the presence of nontrivial spin texture, like skyrmions. Therefore, $\rho_{xy}$ can be expressed as \cite{Huang2012,Soumyanarayanan2017,sivakumar2020topological},

\begin{widetext}
\begin{equation}
 \rho_{xy} = \rho_{OHE} + \rho_{AHE} + \rho_{THE} = R_0 H + R_s M(H) + \rho_{THE} \label{eq:anisotropy}
\end{equation}
\end{widetext}

\begin{figure*}
	\centering
	\includegraphics[width=0.9\linewidth]{"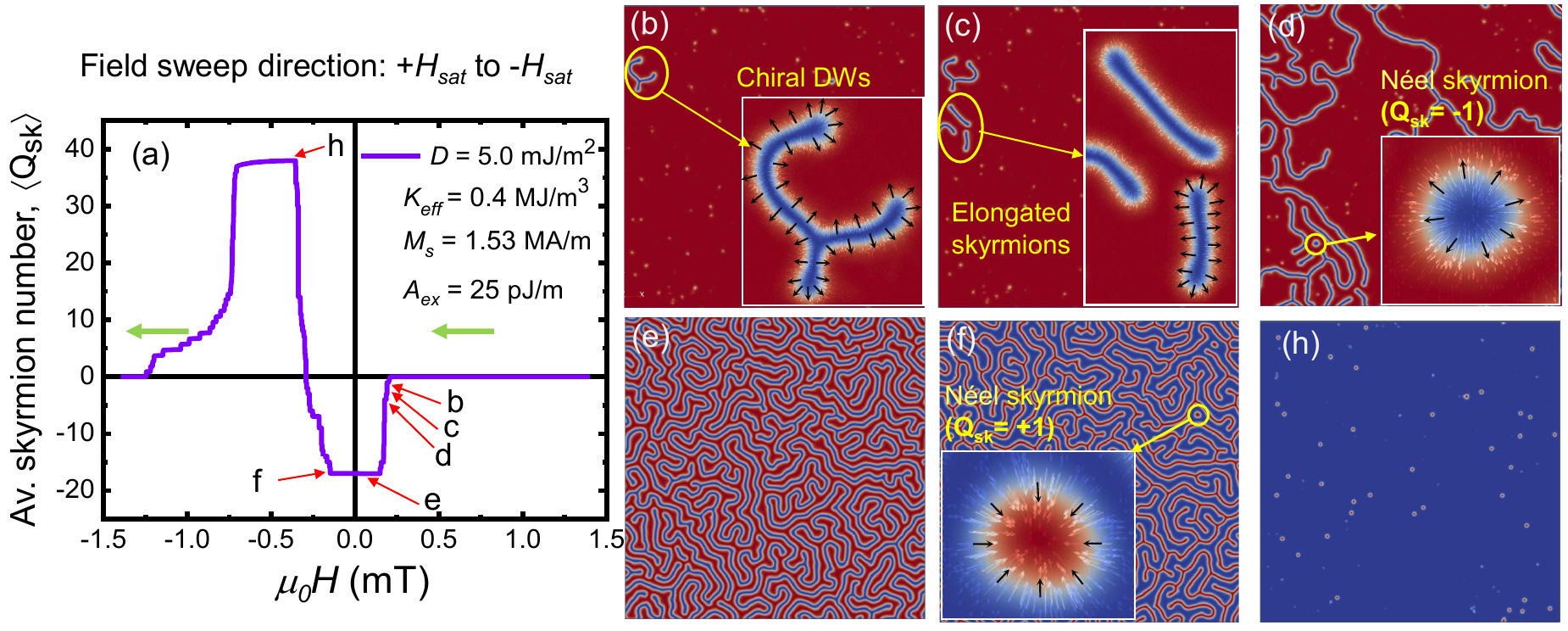"}
	\caption{(a) A double peak with opposite sign in $\left\langle Q_{sk} \right\rangle$ for $D$ = 5.0 mJ/m$^2$ (with $K_{eff}$ = 0.4 MJ/m$^3$, $M_s$=1.53 MA/m, and $A_{ex}$=25 pJ/m) obtained from simulations. The red arrows refer to the corresponding domain structures as shown in (b-g). The green arrows represent the field sweep direction.}
	\label{fig:Fig_6}
\end{figure*}

where $R_0$ and $R_s$ are the ordinary and anomalous Hall coefficients and $M(H)$ is the out-of-plane magnetization as a function of $H$. Thus, $\rho_{THE}$ can be extracted by subtracting the other two components from $\rho_{xy}$. Now, the OHE contribution has been eliminated from $\rho_{xy}$ by fitting a linear function above 1.5 T, and the remaining (AHE + THE) components as a function of $H$ have been shown in Fig. \ref{fig:Fig_2}(a), (c) and (e) for the samples S1, S2 and S3, respectively. At higher fields, magnetization aligns with the field, causing the AHE contribution to saturate. Additionally, due to the absence of chiral objects, $\rho_{THE}$ becomes zero. Therefore, the $R_s$ can be deduced as, $R_s$ =${\rho_{xy, saturation}}/{M_s}$  \cite{Li2019,sivakumar2020topological}. Hence, the AHE components have been determined by scaling the $M-H$ loops with $R_s$ (see Fig. S5 in the supplementary information for details). Now, $\rho_{THE}$ is quantified after subtracting the AHE contributions from the remaining resistivity. The $\rho_{THE}$ as a function of field is shown in Fig. \ref{fig:Fig_2}(b), (d), and (f) for samples S1, S2, and S3, respectively. A maximum $\rho_{THE}$ of $\sim 0.08$, $\sim 0.12$, and $\sim$ 0.10 n$\Omega$ m have been observed for samples S1, S2, and S3, respectively. The topological Hall resistivity confirms the presence of chiral spin textures in the samples. The $\rho_{THE}$ is proportional to the average topological charge or skyrmion number \cite{sivakumar2020topological,Raju2021}:

\begin{equation}
 \rho_{THE} = R_T \left\langle Q_{sk} \right\rangle \label{eq:THE}
\end{equation}

The average skyrmion number, $\left\langle Q_{sk} \right\rangle$, is the sum of the skyrmion numbers, $Q_{sk}$, of all individual objects present in the sample. The topological Hall coefficient, $R_T$, reflects the properties of the band structure, which is governed by the carrier density and the spin polarization of the carriers \cite{sivakumar2020topological,Neubauer2009,raju2019evolution}. The $Q_{sk}$ is defined as a product of vorticity ($V$) and core polarity ($P$) of the texture i.e. $Q_{sk}$ = $V.P$ \cite{sivakumar2020topological,Rohart2013}. The vorticity, $V$ = +1 for skyrmion. The $P$ defines the orientation of a core magnetic moment along the out-of-plane direction of a texture. The value of it is determined by considering the background magnetization. $P$ = +1 (-1) for a -$M_s$ (+$M_s$) magnetized sample. So, depending upon the core polarity, $P$, (+1 or -1), $Q_{sk}$ could be either +1 or -1, which basically defines two distinct phases of topological objects. It should be noted that during the field sweep from +$H_{sat}$ (positive saturation field) to -$H_{sat}$ (negative saturation field) (or vice versa), we have observed a single peak in $\rho_{THE}$ for S1 (Fig. \ref{fig:Fig_2}(b)) which signifies only one type phase of texture. Whereas, a peak and a dip (referred to as a double peak with opposite signs throughout the manuscript) are observed in samples S2 (Fig. \ref{fig:Fig_2}(d)) and S3 (Fig. \ref{fig:Fig_2}(f)) during the field sweep. In the resistivity measurements, the applied field is varied within ±3 T, and the carrier densities do not change significantly with the field. Therefore, the sign of $R_T$ should not change by tuning the applied field. However, $R_T$ may vary with temperature due to modifications in the band structure of the materials and carrier densities. Thus, the sign change in $R_T$ is only possible by changing temperature \cite{sivakumar2020topological}. As all the measurements are performed at room temperature, for the double peak, the sign change in $\rho_{THE}$ solely depends on $Q_{sk}$. During the field sweep, the core polarity switches, resulting in two distinct phases of textures, as discussed later in the manuscript (see Fig. \ref{fig:Fig_6}).

Additionally, from MFM scanning on S1, it has been found that the chiral textures are skyrmions. At a demagnetized state, a worm-like domain state is observed, as shown in Fig. \ref{fig:Fig_3}(a). Further, increasing the field, the worm-like domains start to contract (Fig. \ref{fig:Fig_3}(b)) and form isolated bubbles (skyrmions) as shown in Fig. \ref{fig:Fig_3}(c-d). It is noted that MFM imaging cannot directly reveal the internal chirality of these domains; however, the presence of $\rho_{THE}$ confirms the existence of chiral spin structures. Since all samples exhibit $\rho_{THE}$, the combined analysis of Hall and MFM measurements suggests that the isolated bubbles are skyrmions with a well-defined chirality.  At higher fields, skyrmion density is reduced, and finally, a saturated state is achieved.

Further, the reason behind the different shapes of $\rho_{THE}$ (Fig. \ref{fig:Fig_2}(b) and (d) or (f)) in samples S1 and S2 or S3 has also been understood via micromagnetic simulations. The simulations are performed at 0 K, thereby neglecting thermal fluctuations present in the experimental conditions ($\sim$ 300 K). This leads to some quantitative differences compared to the experimental results. However, the simulations successfully capture the field-dependent evolution of the $\left\langle Q_{sk} \right\rangle$ and  offer a meaningful insight into the mechanisms underlying the single and double peak features observed in the $\rho_{THE}$. Previously, this kind of a single peak \cite{Li2020,Meng2018} and double peak (i.e., double peak with opposite sign) \cite{Soumyanarayanan2017,He2018,Budhathoki2020,Ahmed2019} in resistivity were observed in different thin films. The emergence of a single peak can be attributed to the presence of a singular type of topological object. Conversely, the manifestation of a double peak with opposite signs is indicative of the coexistence of two distinct phases (i.e. $Q_{sk}$ = +1 and -1) of a topological object  \cite{Soumyanarayanan2017,He2018,Budhathoki2020,Ahmed2019}. Recently, P. K. Sivakumar et al. explained that a double peak (with the same sign) in topological resistivity comes due to the presence of skyrmion and anti-skyrmion having the same topological charge in a MgO (001)/Mn$_{2}$RhSn system \cite{sivakumar2020topological}. However, the precise conditions leading to the occurrence of either a single peak or double peak (with opposite signs) in the Hall signature have not been thoroughly investigated. Thus, we have investigated the origin of such behaviour in $\rho_{THE}$ via micromagnetic simulations. In our simulations, the magnetization of the system has been reversed from positive to negative saturation by gradually varying the applied field in small steps. In every field value, the average skyrmion number, $\left\langle Q_{sk} \right\rangle$, is calculated using the ‘$ext-topologicalcharge$’ command in Mumax3. As iDMI defines the chirality of a system, first, the iDMI constant, $D$, is varied in a range of 0.5--5.0 mJ/m$^2$ considering $A_{ex}$ = 25 pJ/m, $M_s$ = 1.53 MA/m and $K_{eff}$ = 0.4 MJ/m$^3$. From Eq. \ref{eq:THE}, it should be noted that $\left\langle Q_{sk} \right\rangle$ corresponds to the $\rho_{THE}$ of the system. Therefore, the simulated $\left\langle Q_{sk} \right\rangle$ successfully reproduces the shape of  $\rho_{THE}$, revealing the origin of single or double peak features.  Fig. \ref{fig:Fig_4}(a) represents the $\left\langle Q_{sk} \right\rangle$ as a function of field. A single peak in $\left\langle Q_{sk} \right\rangle$ has been observed for $D$ = 0.5 to 2.5 mJ/m$^2$. Beyond $D$ = 2.5 mJ/m$^2$, double peak in $\left\langle Q_{sk} \right\rangle$  became prominent. As seen, with enhancing $D$, the maximum value of $\left\langle Q_{sk} \right\rangle$ rises, leading to a transition from a single to a double peak. Thus, $D$ is a very important parameter in deciding the nature $\rho_{THE}$.  

\begin{table*} [t]
\caption{The values of $M_{s}$, $\mu_0 H_s$, and $K_{\text {eff }}$ obtained from SQUID measurements for samples S1-S3.}
	\label{tbl:example}
\begin{tabular}{|l|l|l|l|l|l|}
\hline Sample & $M_s(\mathrm{~MA} / \mathrm{m})$ & $\mu_0 H_s(\mathrm{mT})$ & $K_{\text {eff }}\left(\mathrm{MJ} / \mathrm{m}^3\right)$ & remarks in $D$ & THE signal\\
\hline S1 (Pt/Co(1.5 nm)/Ta)& $1.63 $ & 650 & 0.53 & lower than S2 and S3 & single peak\\
\hline S2 (Pt/Co(1.4 nm)/Re)& $1.51 $ & 720 & 0.54 & higher than S1 & double peak\\
\hline S3 (Pt/Co(1.5 nm)/Re)& $1.53$ & 600 & 0.46 & higher than S1 & double peak \\
\hline
\end{tabular}
\label{table:table1}
\end{table*}

For a single peak in $\left\langle Q_{sk} \right\rangle$ (Fig. \ref{fig:Fig_5}(a), considering $D$ = 2 mJ/m$^2$), while the field is reduced from +$H_{sat}$, the reversed domain with $Q_{sk}$ = 0 nucleates (Fig. \ref{fig:Fig_5}(b)). With further field reduction, this reversed domain expands (Fig. \ref{fig:Fig_5}(c)) and connects with neighboring domains, forming a worm-like structure (Fig. \ref{fig:Fig_5}(d)). As the field continues to decrease (i.e., increase in the negative direction), the worm-like domains break into skyrmions with $ Q_{sk}$ = +1 (Fig. \ref{fig:Fig_5}(e)-(g)), contributing to the topological Hall resistivity. At sufficiently high negative fields, the system transitions back to the oppositely saturated ferromagnetic ground state. Thus, the observed peak in $\left\langle Q_{sk} \right\rangle$ or $\rho_{THE}$ arises solely due to the presence of a single type of topological object.     

In the case of a double peak (Fig. \ref{fig:Fig_6}(a), considering $D$ = 5 mJ/m$^2$), as the field decreases, several reversed nuclei emerge in the form of chiral domain walls (DWs), elongated skyrmions, or skyrmions, contributing to $Q_{sk}$ = -1.  The presence of high $D$ enforces a fixed chirality in the DWs. As the field increases in the reverse direction, these nuclei expand and merge with nearby topological objects, forming a chiral worm-like domain phase, reaching a maximum $\left\langle Q_{sk} \right\rangle$ of -17 (Fig. \ref{fig:Fig_6}(e)). With a further increase in the field, the worm-like phase again breaks into skyrmions of opposite polarity ($Q_{sk}$ = +1, Fig. \ref{fig:Fig_6}(f)), ultimately leading to a state dominated by isolated skyrmions with $\left\langle Q_{sk} \right\rangle$ = +38 (Fig. \ref{fig:Fig_6}(h)). Thus, the double peak in $\left\langle Q_{sk} \right\rangle$ or $\rho_{THE}$ arises from the contributions of two topological objects with opposite polarity.

\begin{figure}[h!]
	\centering
	\includegraphics[width=.9\linewidth] {"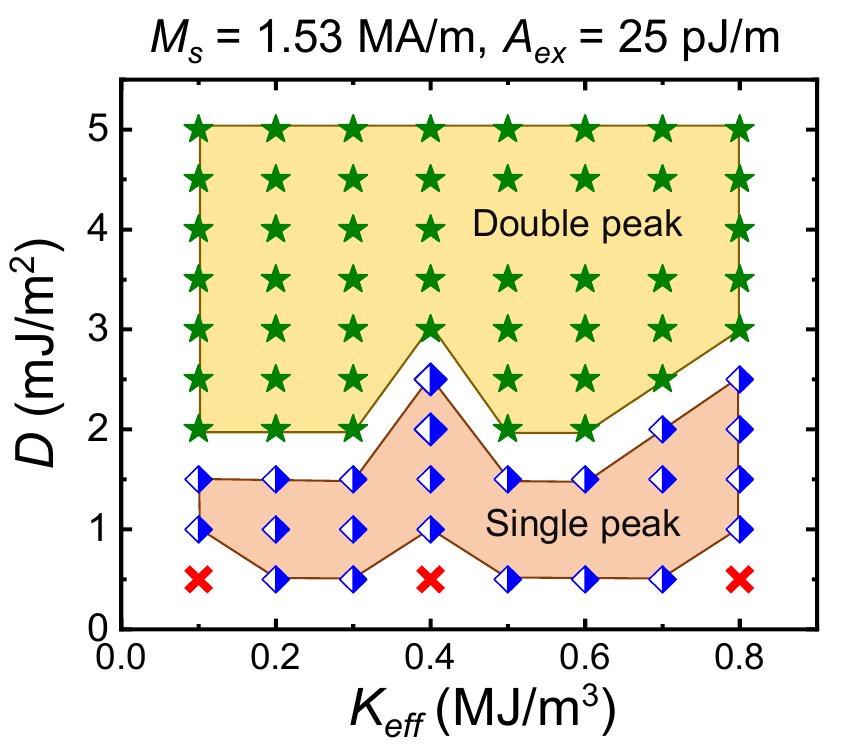"}
\caption{(a) A phase plot of $D$ vs $K_{eff}$ for a constant $A_{ex}$ and $M_s$ which shows the single to double peak transition in $\left\langle Q_{sk} \right\rangle$. The red cross marks represent no skyrmion during the field sweep.}
	\label{fig:Fig_7}
\end{figure}

So far, we have investigated the effect of $D$ on  $\left\langle Q_{sk} \right\rangle$ or $\rho_{THE}$. However, the anisotropy of the system plays a major role in deciding the minimum energy state, which helps to stabilize a chiral texture in a system. The estimated values of $K_{eff}$ are found in the range of 0.45 -- 0.55 MJ/m$^3$, as listed in Table \ref{table:table1}. Prior studies have reported that skyrmion-hosting systems exhibit $K_{eff}$ values in the range of 0.07--0.5 MJ/m$^3$  in experiments \cite{Akhtar2019,Soumyanarayanan2017, ajejas2023densely} and the range of 0.6--1.2 MJ/m$^3$ in simulations \cite{Sampaio2013, chui2015geometrical}. Thus, to systematically investigate the influence of $K_{eff}$ on the THE, simulations are carried out over a broader range of 0.1--0.8 MJ/m$^3$  and found that the shape of the $\left\langle Q_{sk} \right\rangle$ also depends on it. A phase plot of $D$ vs $K_{eff}$ has been shown in Fig. \ref{fig:Fig_7}. 
However, the phase plot is generated for a constant $A_{ex}$ =25 pJ/m and $M_s$ =1.53 MA/m. According to Eq. \ref{eq:energy}, the system's energy also depends on $A_{ex}$ and $M_s$, indicating that the phase diagram is inherently sensitive to variations in these parameters. Consequently, modifying $A_{ex}$ and $M_s$ alters the phase boundaries, leading to significant changes in the phase diagram.

\begin{figure*} 
	\centering
	\includegraphics[width=0.7\linewidth]{"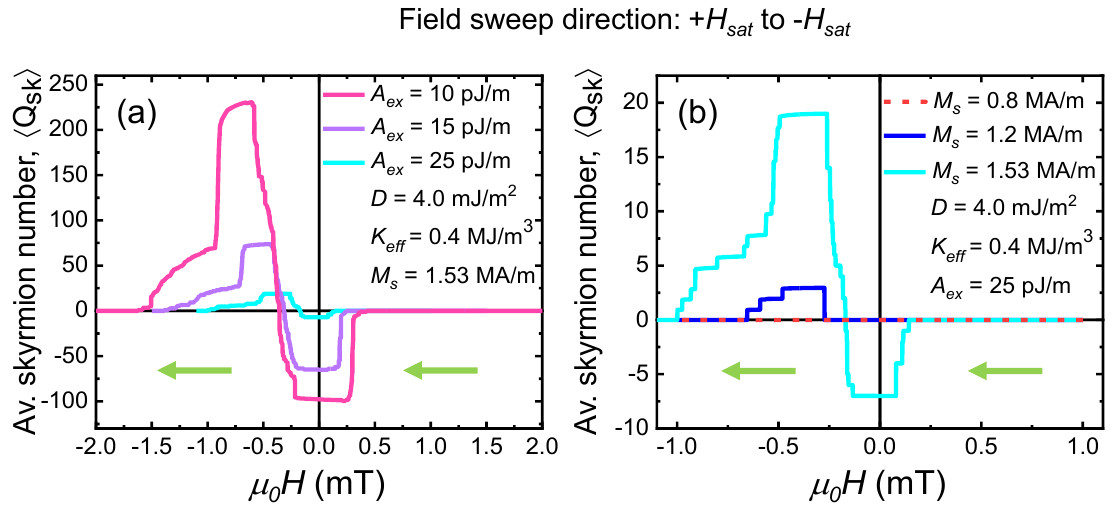"}
	\caption{(a) A double peak with opposite sign in $\left\langle Q_{sk} \right\rangle$ for varying $A_{ex}$ with $D$ = 4.0 mJ/m$^2$,$K_{eff}$ = 0.4 MJ/m$^3$, and $M_s$ = 1.53 MA/m, obtained from simulations. (b) Evolution from a double peak to no peak in $\left\langle Q_{sk} \right\rangle$ for varying $M_s$ with $D$ = 4.0 mJ/m$^2$,$K_{eff}$ = 0.4 MJ/m$^3$, and  $A_{ex}$ = 25 pJ/m. The green arrows represent the field sweep direction.}
	\label{fig:Fig_8}
\end{figure*}

To examine the impact of exchange interaction, we varied $A_{ex}$ in the range of 10--25 pJ/m, consistent with values reported in the literature. In micromagnetic simulations, $A_{ex}$ is typically taken to be in the range of 10--16 pJ/m \cite{Woo2016,Luchaire2016} or 20--25 pJ/m \cite{Davydenko2019,ding2005magnetic}. While experimentally, $A_{ex}$ has been reported in the range of 17--25 pJ/m for Co thin films \cite{Shahbazi2019,vernon1984brillouin,bottcher2023quantifying}.  Therefore, to reflect the differences observed in literature, $A_{ex}$ is varied over the range of 10--25 pJ/m in the simulations. For $A_{ex}$ = 25 pJ/m (with $D$ = 4 mJ/m$^2$, $K_{eff}$ = 0.4 MJ/m$^3$ and $M_s$ = 1.53 MA/m), a double peak is observed in $\left\langle Q_{sk} \right\rangle$, as illustrated in the Fig. \ref{fig:Fig_8}(a). Upon reducing $A_{ex}$ to 15 and 10 pJ/m, the double peak in $\left\langle Q_{sk} \right\rangle$ persists (Fig. \ref{fig:Fig_8}(a)). Notably, this behavior remains unchanged even when reducing $K_{eff}$ (see Fig. S6(a) in supplementary information) or simultaneously lowering both $D$ and $K_{eff}$ (see Fig. S6(b) in supplementary information). The persistence of the double peak can be attributed to the reduced exchange interaction with decreasing  $A_{ex}$, which in turn enhances the relative strength of the iDMI. The strengthen iDMI energy induces a fixed chirality in reversed nuclei domain, leading to $Q_{sk}$ = -1 as the field decreases from +$H_{sat}$ (similar to Fig. \ref{fig:Fig_6}(b-d)). Further field reduction leads to a transition from chiral worm-like domains to skyrmions, resulting in $Q_{sk}$ =+1 (similar to Fig. \ref{fig:Fig_6}(e-h)) and thereby sustaining the double peak structure with decreasing $A_{ex}$.

Additionally, $M_s$ is varied within the range of 0.8--1.53 MA/m, revealing a transition from a double peak to a single peak in $\left\langle Q_{sk} \right\rangle$ as $M_s$ decreases. For $M_s$=1.53 MA/m (with $D$=4 mJ/m², $K_{eff}$=0.4 MJ/m³, and $A_{ex}$=25 pJ/m), a double peak is observed in $\left\langle Q_{sk} \right\rangle$, as shown in  Fig. \ref{fig:Fig_8}(b). However, reducing $M_s$ to 1.2 MA/m results in a single peak (see Fig. \ref{fig:Fig_8}(b)). A further decrease to 0.8 MA/m leads to the complete suppression of the peak in $\left\langle Q_{sk} \right\rangle$ (Fig. \ref{fig:Fig_8}(b)), demonstrating the strong dependence of $\left\langle Q_{sk} \right\rangle$ on $M_s$. Further, we varied both $D$ and $K_{eff}$ while tuning the $M_s$ (see Fig. S7 in supplementary information). Reducing $K_{eff}$ results in a transition in 
$\left\langle Q_{sk} \right\rangle$ from a double peak to a single peak and eventually to no peak, as $M_s$ decreases (see Fig. S7(a) in supplementary information).  Reducing $D$ to 1 mJ/m², a single peak is observed for $M_s$ = 1.53 MA/m, while further reduction in $M_s$ results in the suppression of the peak (see Fig. S7(b)-(c) in supplementary information). These observations indicate that the number of peaks in $\left\langle Q_{sk} \right\rangle$ consistently decreases or vanishes as $M_s$ decreases for a given $D$. This behavior can be attributed to the reduction of shape anisotropy with decreasing the $M_s$, which corresponds to a decrease in in-plane anisotropy, leading to an enhancement in OOP anisotropy. The increased OOP anisotropy suppresses the formation of a fixed chirality in the reversed nuclei domain  (similar to Fig. \ref{fig:Fig_5}(b)), ultimately leading to a reduction in the number of peaks as $M_s$ decreases.

As observed, the shape of  $\left\langle Q_{sk} \right\rangle$  or $\rho_{THE}$ governed by the magnetic parameters $A_{ex}$, $D$, $K_{eff}$, and $M_s$. By carefully tuning these parameters, a transition from a single to double peak in $\rho_{THE}$ can be achieved. In all investigated thin films, $K_{eff}$, and $M_s$ remain nearly identical. Moreover, since Co is used as the ferromagnetic layer in all samples, significant variations in $A_{ex}$ are unlikely. However, the value of $D$ is expected to be lower in the Pt/Co/Ta sample (S1) compare to the Pt/Co/Re  samples (S2 and S3) \cite{Shahbazi2019,Kashid2014,Jadaun12020,fakhredine2024huge,Nomura2022}. Thus, the analysis suggests that the relatively lower iDMI in Pt/Co/Ta results a single peak in $\rho_{THE}$, whereas the higher iDMI in Pt/Co/Re leads to the emergence of a double peak.

\section{Conclusions}
In summary, the THE has been observed in both Pt/Co/Ta and Pt/Co/Re systems, indicating the existence of skyrmions, which is confirmed via MFM imaging. A maximum $\rho_{THE}$ of $\sim$0.08 and $\sim$0.12 n$\Omega$.m has been observed for sample Pt/Co/Ta and Pt/Co/Re, respectively. Two different natures of topological Hall resistivity are observed in the samples. As seen, the $A_{ex}$, $D$, $K_{eff}$, and $M_s$ play a pivotal role in determining THE signal in a system. By choosing a proper combination of these parameters of a system, one can achieve single to double peak transition. The presence of a single peak and double peak with opposite signs in THE signal corresponds to the stabilization of one or two distinct skyrmionic phases during magnetization reversal. Furthermore, tuning the $A_{ex}$, $D$, $K_{eff}$, and $M_s$ values may provide a pathway for the design of novel material systems that stabilize multiple skyrmionic phases. We believe that such control over skyrmion phases could bring additional functionality to skyrmion-based memory \cite{tomasello2014strategy,luo2021skyrmion}, logic devices \cite{zhang2015magnetic}, or reservoir computing performance \cite{raab2022brownian}. 

\section{Acknowledgments}
The authors would like to thank Dr. Esita Pandey for the valuable discussion and help during manuscript preparation. The authors also thank the Department of Atomic Energy (DAE) of India, the Indo-French collaborative project supported by CEFIPRA (IFC/5808-1/2017), and DST-SERB (Grant No. CRG/2021/001245) for financial support.

\bibliography{References} 

\begin{thebibliography}{70}%
\makeatletter
\providecommand \@ifxundefined [1]{%
 \@ifx{#1\undefined}
}%
\providecommand \@ifnum [1]{%
 \ifnum #1\expandafter \@firstoftwo
 \else \expandafter \@secondoftwo
 \fi
}%
\providecommand \@ifx [1]{%
 \ifx #1\expandafter \@firstoftwo
 \else \expandafter \@secondoftwo
 \fi
}%
\providecommand \natexlab [1]{#1}%
\providecommand \enquote  [1]{``#1''}%
\providecommand \bibnamefont  [1]{#1}%
\providecommand \bibfnamefont [1]{#1}%
\providecommand \citenamefont [1]{#1}%
\providecommand \href@noop [0]{\@secondoftwo}%
\providecommand \href [0]{\begingroup \@sanitize@url \@href}%
\providecommand \@href[1]{\@@startlink{#1}\@@href}%
\providecommand \@@href[1]{\endgroup#1\@@endlink}%
\providecommand \@sanitize@url [0]{\catcode `\\12\catcode `\$12\catcode `\&12\catcode `\#12\catcode `\^12\catcode `\_12\catcode `\%12\relax}%
\providecommand \@@startlink[1]{}%
\providecommand \@@endlink[0]{}%
\providecommand \url  [0]{\begingroup\@sanitize@url \@url }%
\providecommand \@url [1]{\endgroup\@href {#1}{\urlprefix }}%
\providecommand \urlprefix  [0]{URL }%
\providecommand \Eprint [0]{\href }%
\providecommand \doibase [0]{https://doi.org/}%
\providecommand \selectlanguage [0]{\@gobble}%
\providecommand \bibinfo  [0]{\@secondoftwo}%
\providecommand \bibfield  [0]{\@secondoftwo}%
\providecommand \translation [1]{[#1]}%
\providecommand \BibitemOpen [0]{}%
\providecommand \bibitemStop [0]{}%
\providecommand \bibitemNoStop [0]{.\EOS\space}%
\providecommand \EOS [0]{\spacefactor3000\relax}%
\providecommand \BibitemShut  [1]{\csname bibitem#1\endcsname}%
\let\auto@bib@innerbib\@empty
\bibitem [{\citenamefont {Fert}\ \emph {et~al.}(2017)\citenamefont {Fert}, \citenamefont {Reyren},\ and\ \citenamefont {Cros}}]{Fert2017}%
  \BibitemOpen
  \bibfield  {author} {\bibinfo {author} {\bibfnamefont {A.}~\bibnamefont {Fert}}, \bibinfo {author} {\bibfnamefont {N.}~\bibnamefont {Reyren}},\ and\ \bibinfo {author} {\bibfnamefont {V.}~\bibnamefont {Cros}},\ }\bibfield  {title} {\bibinfo {title} {Magnetic skyrmions: advances in physics and potential applications},\ }\href {https://doi.org/10.1038/natrevmats.2017.31} {\bibfield  {journal} {\bibinfo  {journal} {Nature Reviews Materials}\ }\textbf {\bibinfo {volume} {2}},\ \bibinfo {pages} {17031} (\bibinfo {year} {2017})}\BibitemShut {NoStop}%
\bibitem [{\citenamefont {Everschor-Sitte}\ \emph {et~al.}(2018)\citenamefont {Everschor-Sitte}, \citenamefont {Masell}, \citenamefont {Reeve},\ and\ \citenamefont {Kläui}}]{Karin2018}%
  \BibitemOpen
  \bibfield  {author} {\bibinfo {author} {\bibfnamefont {K.}~\bibnamefont {Everschor-Sitte}}, \bibinfo {author} {\bibfnamefont {J.}~\bibnamefont {Masell}}, \bibinfo {author} {\bibfnamefont {R.~M.}\ \bibnamefont {Reeve}},\ and\ \bibinfo {author} {\bibfnamefont {M.}~\bibnamefont {Kläui}},\ }\bibfield  {title} {\bibinfo {title} {Perspective: Magnetic skyrmions--overview of recent progress in an active research field},\ }\href {https://doi.org/10.1063/1.5048972} {\bibfield  {journal} {\bibinfo  {journal} {Journal of Applied Physics}\ }\textbf {\bibinfo {volume} {124}},\ \bibinfo {pages} {240901} (\bibinfo {year} {2018})}\BibitemShut {NoStop}%
\bibitem [{\citenamefont {Fert}\ \emph {et~al.}(2013)\citenamefont {Fert}, \citenamefont {Cros},\ and\ \citenamefont {Sampaio}}]{Fert2013}%
  \BibitemOpen
  \bibfield  {author} {\bibinfo {author} {\bibfnamefont {A.}~\bibnamefont {Fert}}, \bibinfo {author} {\bibfnamefont {V.}~\bibnamefont {Cros}},\ and\ \bibinfo {author} {\bibfnamefont {J.}~\bibnamefont {Sampaio}},\ }\bibfield  {title} {\bibinfo {title} {Skyrmions on the track},\ }\href {https://doi.org/10.1038/nnano.2013.29} {\bibfield  {journal} {\bibinfo  {journal} {Nature Nanotechnology}\ }\textbf {\bibinfo {volume} {8}},\ \bibinfo {pages} {152} (\bibinfo {year} {2013})}\BibitemShut {NoStop}%
\bibitem [{\citenamefont {Luo}\ and\ \citenamefont {You}(2021{\natexlab{a}})}]{Luo2021}%
  \BibitemOpen
  \bibfield  {author} {\bibinfo {author} {\bibfnamefont {S.}~\bibnamefont {Luo}}\ and\ \bibinfo {author} {\bibfnamefont {L.}~\bibnamefont {You}},\ }\bibfield  {title} {\bibinfo {title} {Skyrmion devices for memory and logic applications},\ }\href {https://doi.org/10.1063/5.0042917} {\bibfield  {journal} {\bibinfo  {journal} {APL Materials}\ }\textbf {\bibinfo {volume} {9}},\ \bibinfo {pages} {050901} (\bibinfo {year} {2021}{\natexlab{a}})}\BibitemShut {NoStop}%
\bibitem [{\citenamefont {Tomasello}\ \emph {et~al.}(2014{\natexlab{a}})\citenamefont {Tomasello}, \citenamefont {Martinez}, \citenamefont {Zivieri}, \citenamefont {Torres}, \citenamefont {Carpentieri},\ and\ \citenamefont {Finocchio}}]{Tomasello2014}%
  \BibitemOpen
  \bibfield  {author} {\bibinfo {author} {\bibfnamefont {R.}~\bibnamefont {Tomasello}}, \bibinfo {author} {\bibfnamefont {E.}~\bibnamefont {Martinez}}, \bibinfo {author} {\bibfnamefont {R.}~\bibnamefont {Zivieri}}, \bibinfo {author} {\bibfnamefont {L.}~\bibnamefont {Torres}}, \bibinfo {author} {\bibfnamefont {M.}~\bibnamefont {Carpentieri}},\ and\ \bibinfo {author} {\bibfnamefont {G.}~\bibnamefont {Finocchio}},\ }\bibfield  {title} {\bibinfo {title} {A strategy for the design of skyrmion racetrack memories},\ }\href {https://doi.org/10.1038/srep06784} {\bibfield  {journal} {\bibinfo  {journal} {Scientific Reports}\ }\textbf {\bibinfo {volume} {4}},\ \bibinfo {pages} {6784} (\bibinfo {year} {2014}{\natexlab{a}})}\BibitemShut {NoStop}%
\bibitem [{\citenamefont {Jiang}\ \emph {et~al.}(2015)\citenamefont {Jiang}, \citenamefont {Upadhyaya}, \citenamefont {Zhang}, \citenamefont {Yu}, \citenamefont {Jungfleisch}, \citenamefont {Fradin}, \citenamefont {Pearson}, \citenamefont {Tserkovnyak}, \citenamefont {Wang}, \citenamefont {Heinonen} \emph {et~al.}}]{Jiang2015}%
  \BibitemOpen
  \bibfield  {author} {\bibinfo {author} {\bibfnamefont {W.}~\bibnamefont {Jiang}}, \bibinfo {author} {\bibfnamefont {P.}~\bibnamefont {Upadhyaya}}, \bibinfo {author} {\bibfnamefont {W.}~\bibnamefont {Zhang}}, \bibinfo {author} {\bibfnamefont {G.}~\bibnamefont {Yu}}, \bibinfo {author} {\bibfnamefont {M.~B.}\ \bibnamefont {Jungfleisch}}, \bibinfo {author} {\bibfnamefont {F.~Y.}\ \bibnamefont {Fradin}}, \bibinfo {author} {\bibfnamefont {J.~E.}\ \bibnamefont {Pearson}}, \bibinfo {author} {\bibfnamefont {Y.}~\bibnamefont {Tserkovnyak}}, \bibinfo {author} {\bibfnamefont {K.~L.}\ \bibnamefont {Wang}}, \bibinfo {author} {\bibfnamefont {O.}~\bibnamefont {Heinonen}}, \emph {et~al.},\ }\bibfield  {title} {\bibinfo {title} {Blowing magnetic skyrmion bubbles},\ }\href {https://doi.org/10.1126/science.aaa1442} {\bibfield  {journal} {\bibinfo  {journal} {Science}\ }\textbf {\bibinfo {volume} {349}},\ \bibinfo {pages} {283} (\bibinfo {year} {2015})}\BibitemShut {NoStop}%
\bibitem [{\citenamefont {Moreau-Luchaire}\ \emph {et~al.}(2016)\citenamefont {Moreau-Luchaire}, \citenamefont {Moutafis}, \citenamefont {Reyren}, \citenamefont {Sampaio}, \citenamefont {Vaz}, \citenamefont {Horne}, \citenamefont {Bouzehouane}, \citenamefont {Garcia}, \citenamefont {Deranlot}, \citenamefont {Warnicke} \emph {et~al.}}]{Luchaire2016}%
  \BibitemOpen
  \bibfield  {author} {\bibinfo {author} {\bibfnamefont {C.}~\bibnamefont {Moreau-Luchaire}}, \bibinfo {author} {\bibfnamefont {C.}~\bibnamefont {Moutafis}}, \bibinfo {author} {\bibfnamefont {N.}~\bibnamefont {Reyren}}, \bibinfo {author} {\bibfnamefont {J.}~\bibnamefont {Sampaio}}, \bibinfo {author} {\bibfnamefont {C.~A.~F.}\ \bibnamefont {Vaz}}, \bibinfo {author} {\bibfnamefont {N.~V.}\ \bibnamefont {Horne}}, \bibinfo {author} {\bibfnamefont {K.}~\bibnamefont {Bouzehouane}}, \bibinfo {author} {\bibfnamefont {K.}~\bibnamefont {Garcia}}, \bibinfo {author} {\bibfnamefont {C.}~\bibnamefont {Deranlot}}, \bibinfo {author} {\bibfnamefont {P.}~\bibnamefont {Warnicke}}, \emph {et~al.},\ }\bibfield  {title} {\bibinfo {title} {Additive interfacial chiral interaction in multilayers for stabilization of small individual skyrmions at room temperature},\ }\href {https://doi.org/10.1038/nnano.2015.313} {\bibfield  {journal} {\bibinfo  {journal} {Nature Nanotechnology}\ }\textbf {\bibinfo {volume} {11}},\ \bibinfo {pages}
  {444} (\bibinfo {year} {2016})}\BibitemShut {NoStop}%
\bibitem [{\citenamefont {Hrabec}\ \emph {et~al.}(2017)\citenamefont {Hrabec}, \citenamefont {Sampaio}, \citenamefont {Belmeguenai}, \citenamefont {Gross}, \citenamefont {Weil}, \citenamefont {Ch{\'e}rif}, \citenamefont {Stashkevich}, \citenamefont {Jacques}, \citenamefont {Thiaville},\ and\ \citenamefont {Rohart}}]{Hrabec2017}%
  \BibitemOpen
  \bibfield  {author} {\bibinfo {author} {\bibfnamefont {A.}~\bibnamefont {Hrabec}}, \bibinfo {author} {\bibfnamefont {J.}~\bibnamefont {Sampaio}}, \bibinfo {author} {\bibfnamefont {M.}~\bibnamefont {Belmeguenai}}, \bibinfo {author} {\bibfnamefont {I.}~\bibnamefont {Gross}}, \bibinfo {author} {\bibfnamefont {R.}~\bibnamefont {Weil}}, \bibinfo {author} {\bibfnamefont {S.~M.}\ \bibnamefont {Ch{\'e}rif}}, \bibinfo {author} {\bibfnamefont {A.}~\bibnamefont {Stashkevich}}, \bibinfo {author} {\bibfnamefont {V.}~\bibnamefont {Jacques}}, \bibinfo {author} {\bibfnamefont {A.}~\bibnamefont {Thiaville}},\ and\ \bibinfo {author} {\bibfnamefont {S.}~\bibnamefont {Rohart}},\ }\bibfield  {title} {\bibinfo {title} {Current-induced skyrmion generation and dynamics in symmetric bilayers},\ }\href {https://doi.org/10.1038/ncomms15765} {\bibfield  {journal} {\bibinfo  {journal} {Nature Communications}\ }\textbf {\bibinfo {volume} {8}},\ \bibinfo {pages} {15765} (\bibinfo {year} {2017})}\BibitemShut {NoStop}%
\bibitem [{\citenamefont {Ojha}\ \emph {et~al.}(2023)\citenamefont {Ojha}, \citenamefont {Mallick}, \citenamefont {Panigrahy}, \citenamefont {Sharma}, \citenamefont {Thiaville}, \citenamefont {Rohart},\ and\ \citenamefont {Bedanta}}]{Ojha2023}%
  \BibitemOpen
  \bibfield  {author} {\bibinfo {author} {\bibfnamefont {B.}~\bibnamefont {Ojha}}, \bibinfo {author} {\bibfnamefont {S.}~\bibnamefont {Mallick}}, \bibinfo {author} {\bibfnamefont {S.}~\bibnamefont {Panigrahy}}, \bibinfo {author} {\bibfnamefont {M.}~\bibnamefont {Sharma}}, \bibinfo {author} {\bibfnamefont {A.}~\bibnamefont {Thiaville}}, \bibinfo {author} {\bibfnamefont {S.}~\bibnamefont {Rohart}},\ and\ \bibinfo {author} {\bibfnamefont {S.}~\bibnamefont {Bedanta}},\ }\bibfield  {title} {\bibinfo {title} {Driving skyrmions with low threshold current density in $\mathrm{Pt}/\mathrm{CoFeB}$ thin film},\ }\href {https://dx.doi.org/10.1088/1402-4896/acb862} {\bibfield  {journal} {\bibinfo  {journal} {Physica Scripta}\ }\textbf {\bibinfo {volume} {98}},\ \bibinfo {pages} {035819} (\bibinfo {year} {2023})}\BibitemShut {NoStop}%
\bibitem [{\citenamefont {Bogdanov}\ and\ \citenamefont {Hubert}(1994)}]{bogdanov1994thermodynamically}%
  \BibitemOpen
  \bibfield  {author} {\bibinfo {author} {\bibfnamefont {A.}~\bibnamefont {Bogdanov}}\ and\ \bibinfo {author} {\bibfnamefont {A.}~\bibnamefont {Hubert}},\ }\bibfield  {title} {\bibinfo {title} {Thermodynamically stable magnetic vortex states in magnetic crystals},\ }\href {https://doi.org/10.1016/0304-8853(94)90046-9} {\bibfield  {journal} {\bibinfo  {journal} {Journal of Magnetism and Magnetic Materials}\ }\textbf {\bibinfo {volume} {138}},\ \bibinfo {pages} {255} (\bibinfo {year} {1994})}\BibitemShut {NoStop}%
\bibitem [{\citenamefont {Bogdanov}\ and\ \citenamefont {Panagopoulos}(2020)}]{bogdanov2020physical}%
  \BibitemOpen
  \bibfield  {author} {\bibinfo {author} {\bibfnamefont {A.~N.}\ \bibnamefont {Bogdanov}}\ and\ \bibinfo {author} {\bibfnamefont {C.}~\bibnamefont {Panagopoulos}},\ }\bibfield  {title} {\bibinfo {title} {Physical foundations and basic properties of magnetic skyrmions},\ }\href {https://doi.org/10.1038/s42254-020-0203-7} {\bibfield  {journal} {\bibinfo  {journal} {Nature Reviews Physics}\ }\textbf {\bibinfo {volume} {2}},\ \bibinfo {pages} {492} (\bibinfo {year} {2020})}\BibitemShut {NoStop}%
\bibitem [{\citenamefont {Romming}\ \emph {et~al.}(2015)\citenamefont {Romming}, \citenamefont {Kubetzka}, \citenamefont {Hanneken}, \citenamefont {von Bergmann},\ and\ \citenamefont {Wiesendanger}}]{Romming2015}%
  \BibitemOpen
  \bibfield  {author} {\bibinfo {author} {\bibfnamefont {N.}~\bibnamefont {Romming}}, \bibinfo {author} {\bibfnamefont {A.}~\bibnamefont {Kubetzka}}, \bibinfo {author} {\bibfnamefont {C.}~\bibnamefont {Hanneken}}, \bibinfo {author} {\bibfnamefont {K.}~\bibnamefont {von Bergmann}},\ and\ \bibinfo {author} {\bibfnamefont {R.}~\bibnamefont {Wiesendanger}},\ }\bibfield  {title} {\bibinfo {title} {Field-dependent size and shape of single magnetic skyrmions},\ }\href {https://doi.org/10.1103/PhysRevLett.114.177203} {\bibfield  {journal} {\bibinfo  {journal} {Physical Review Letters}\ }\textbf {\bibinfo {volume} {114}},\ \bibinfo {pages} {177203} (\bibinfo {year} {2015})}\BibitemShut {NoStop}%
\bibitem [{\citenamefont {Heinze}\ \emph {et~al.}(2011)\citenamefont {Heinze}, \citenamefont {Bergmann}, \citenamefont {Menzel}, \citenamefont {Brede}, \citenamefont {Kubetzka}, \citenamefont {Wiesendanger}, \citenamefont {Bihlmayer},\ and\ \citenamefont {Blügel}}]{Heinze2011}%
  \BibitemOpen
  \bibfield  {author} {\bibinfo {author} {\bibfnamefont {S.}~\bibnamefont {Heinze}}, \bibinfo {author} {\bibfnamefont {K.~V.}\ \bibnamefont {Bergmann}}, \bibinfo {author} {\bibfnamefont {M.}~\bibnamefont {Menzel}}, \bibinfo {author} {\bibfnamefont {J.}~\bibnamefont {Brede}}, \bibinfo {author} {\bibfnamefont {A.}~\bibnamefont {Kubetzka}}, \bibinfo {author} {\bibfnamefont {R.}~\bibnamefont {Wiesendanger}}, \bibinfo {author} {\bibfnamefont {G.}~\bibnamefont {Bihlmayer}},\ and\ \bibinfo {author} {\bibfnamefont {S.}~\bibnamefont {Blügel}},\ }\bibfield  {title} {\bibinfo {title} {Spontaneous atomic-scale magnetic skyrmion lattice in two dimensions},\ }\href {https://doi.org/10.1038/nphys2045} {\bibfield  {journal} {\bibinfo  {journal} {Nature Physics}\ }\textbf {\bibinfo {volume} {7}},\ \bibinfo {pages} {713} (\bibinfo {year} {2011})}\BibitemShut {NoStop}%
\bibitem [{\citenamefont {Woo}\ \emph {et~al.}(2018)\citenamefont {Woo}, \citenamefont {Song}, \citenamefont {Zhang}, \citenamefont {Zhou}, \citenamefont {Ezawa}, \citenamefont {Liu}, \citenamefont {Finizio}, \citenamefont {Raabe}, \citenamefont {Lee}, \citenamefont {Kim} \emph {et~al.}}]{Woo2018}%
  \BibitemOpen
  \bibfield  {author} {\bibinfo {author} {\bibfnamefont {S.}~\bibnamefont {Woo}}, \bibinfo {author} {\bibfnamefont {K.~M.}\ \bibnamefont {Song}}, \bibinfo {author} {\bibfnamefont {X.}~\bibnamefont {Zhang}}, \bibinfo {author} {\bibfnamefont {Y.}~\bibnamefont {Zhou}}, \bibinfo {author} {\bibfnamefont {M.}~\bibnamefont {Ezawa}}, \bibinfo {author} {\bibfnamefont {X.}~\bibnamefont {Liu}}, \bibinfo {author} {\bibfnamefont {S.}~\bibnamefont {Finizio}}, \bibinfo {author} {\bibfnamefont {J.}~\bibnamefont {Raabe}}, \bibinfo {author} {\bibfnamefont {N.~J.}\ \bibnamefont {Lee}}, \bibinfo {author} {\bibfnamefont {S.-I.}\ \bibnamefont {Kim}}, \emph {et~al.},\ }\bibfield  {title} {\bibinfo {title} {Current-driven dynamics and inhibition of the skyrmion $\mathrm{Hall}$ effect of ferrimagnetic skyrmions in $\mathrm{GdFeCo}$ films},\ }\href {https://doi.org/10.1038/s41467-018-03378-7} {\bibfield  {journal} {\bibinfo  {journal} {Nature Communications}\ }\textbf {\bibinfo {volume} {9}},\ \bibinfo {pages} {959} (\bibinfo {year}
  {2018})}\BibitemShut {NoStop}%
\bibitem [{\citenamefont {Woo}\ \emph {et~al.}(2016)\citenamefont {Woo}, \citenamefont {Litzius}, \citenamefont {Krüger}, \citenamefont {Im}, \citenamefont {Caretta}, \citenamefont {Richter}, \citenamefont {Mann}, \citenamefont {Krone}, \citenamefont {Reeve}, \citenamefont {Weigand} \emph {et~al.}}]{Woo2016}%
  \BibitemOpen
  \bibfield  {author} {\bibinfo {author} {\bibfnamefont {S.}~\bibnamefont {Woo}}, \bibinfo {author} {\bibfnamefont {K.}~\bibnamefont {Litzius}}, \bibinfo {author} {\bibfnamefont {B.}~\bibnamefont {Krüger}}, \bibinfo {author} {\bibfnamefont {M.-Y.}\ \bibnamefont {Im}}, \bibinfo {author} {\bibfnamefont {L.}~\bibnamefont {Caretta}}, \bibinfo {author} {\bibfnamefont {K.}~\bibnamefont {Richter}}, \bibinfo {author} {\bibfnamefont {M.}~\bibnamefont {Mann}}, \bibinfo {author} {\bibfnamefont {A.}~\bibnamefont {Krone}}, \bibinfo {author} {\bibfnamefont {R.~M.}\ \bibnamefont {Reeve}}, \bibinfo {author} {\bibfnamefont {M.}~\bibnamefont {Weigand}}, \emph {et~al.},\ }\bibfield  {title} {\bibinfo {title} {Observation of room-temperature magnetic skyrmions and their current-driven dynamics in ultrathin metallic ferromagnets},\ }\href {https://doi.org/10.1038/nmat4593} {\bibfield  {journal} {\bibinfo  {journal} {Nature Materials}\ }\textbf {\bibinfo {volume} {15}},\ \bibinfo {pages} {501} (\bibinfo {year}
  {2016})}\BibitemShut {NoStop}%
\bibitem [{\citenamefont {Legrand}\ \emph {et~al.}(2020)\citenamefont {Legrand}, \citenamefont {Maccariello}, \citenamefont {Ajejas}, \citenamefont {Collin}, \citenamefont {Vecchiola}, \citenamefont {Bouzehouane}, \citenamefont {Reyren}, \citenamefont {Cros},\ and\ \citenamefont {Fert}}]{Legrand2020}%
  \BibitemOpen
  \bibfield  {author} {\bibinfo {author} {\bibfnamefont {W.}~\bibnamefont {Legrand}}, \bibinfo {author} {\bibfnamefont {D.}~\bibnamefont {Maccariello}}, \bibinfo {author} {\bibfnamefont {F.}~\bibnamefont {Ajejas}}, \bibinfo {author} {\bibfnamefont {S.}~\bibnamefont {Collin}}, \bibinfo {author} {\bibfnamefont {A.}~\bibnamefont {Vecchiola}}, \bibinfo {author} {\bibfnamefont {K.}~\bibnamefont {Bouzehouane}}, \bibinfo {author} {\bibfnamefont {N.}~\bibnamefont {Reyren}}, \bibinfo {author} {\bibfnamefont {V.}~\bibnamefont {Cros}},\ and\ \bibinfo {author} {\bibfnamefont {A.}~\bibnamefont {Fert}},\ }\bibfield  {title} {\bibinfo {title} {Room-temperature stabilization of antiferromagnetic skyrmions in synthetic antiferromagnets},\ }\href {https://doi.org/10.1038/s41563-019-0468-3} {\bibfield  {journal} {\bibinfo  {journal} {Nature Materials}\ }\textbf {\bibinfo {volume} {19}},\ \bibinfo {pages} {34} (\bibinfo {year} {2020})}\BibitemShut {NoStop}%
\bibitem [{\citenamefont {Mallick}\ \emph {et~al.}(2024)\citenamefont {Mallick}, \citenamefont {Sassi}, \citenamefont {Prestes}, \citenamefont {Krishnia}, \citenamefont {Gallego}, \citenamefont {M.~Vicente~Arche}, \citenamefont {Denneulin}, \citenamefont {Collin}, \citenamefont {Bouzehouane}, \citenamefont {Thiaville} \emph {et~al.}}]{mallick2024driving}%
  \BibitemOpen
  \bibfield  {author} {\bibinfo {author} {\bibfnamefont {S.}~\bibnamefont {Mallick}}, \bibinfo {author} {\bibfnamefont {Y.}~\bibnamefont {Sassi}}, \bibinfo {author} {\bibfnamefont {N.~F.}\ \bibnamefont {Prestes}}, \bibinfo {author} {\bibfnamefont {S.}~\bibnamefont {Krishnia}}, \bibinfo {author} {\bibfnamefont {F.}~\bibnamefont {Gallego}}, \bibinfo {author} {\bibfnamefont {L.}~\bibnamefont {M.~Vicente~Arche}}, \bibinfo {author} {\bibfnamefont {T.}~\bibnamefont {Denneulin}}, \bibinfo {author} {\bibfnamefont {S.}~\bibnamefont {Collin}}, \bibinfo {author} {\bibfnamefont {K.}~\bibnamefont {Bouzehouane}}, \bibinfo {author} {\bibfnamefont {A.}~\bibnamefont {Thiaville}}, \emph {et~al.},\ }\bibfield  {title} {\bibinfo {title} {Driving skyrmions in flow regime in synthetic ferrimagnets},\ }\href {https://doi.org/10.1038/s41467-024-52210-y} {\bibfield  {journal} {\bibinfo  {journal} {Nature Communications}\ }\textbf {\bibinfo {volume} {15}},\ \bibinfo {pages} {8472} (\bibinfo {year} {2024})}\BibitemShut {NoStop}%
\bibitem [{\citenamefont {Wu}\ \emph {et~al.}(2020)\citenamefont {Wu}, \citenamefont {Zhang}, \citenamefont {Zhang}, \citenamefont {Wang}, \citenamefont {Zhu}, \citenamefont {Hu}, \citenamefont {Yin}, \citenamefont {Wong}, \citenamefont {Fang}, \citenamefont {Wan} \emph {et~al.}}]{Wu2020}%
  \BibitemOpen
  \bibfield  {author} {\bibinfo {author} {\bibfnamefont {Y.}~\bibnamefont {Wu}}, \bibinfo {author} {\bibfnamefont {S.}~\bibnamefont {Zhang}}, \bibinfo {author} {\bibfnamefont {J.}~\bibnamefont {Zhang}}, \bibinfo {author} {\bibfnamefont {W.}~\bibnamefont {Wang}}, \bibinfo {author} {\bibfnamefont {Y.~L.}\ \bibnamefont {Zhu}}, \bibinfo {author} {\bibfnamefont {J.}~\bibnamefont {Hu}}, \bibinfo {author} {\bibfnamefont {G.}~\bibnamefont {Yin}}, \bibinfo {author} {\bibfnamefont {K.}~\bibnamefont {Wong}}, \bibinfo {author} {\bibfnamefont {C.}~\bibnamefont {Fang}}, \bibinfo {author} {\bibfnamefont {C.}~\bibnamefont {Wan}}, \emph {et~al.},\ }\bibfield  {title} {\bibinfo {title} {Néel-type skyrmion in $\mathrm{WTe_{2}}/\mathrm{Fe_{3}GeTe_{2}}$ van der waals heterostructure},\ }\href {https://doi.org/10.1038/s41467-020-17566-x} {\bibfield  {journal} {\bibinfo  {journal} {Nature Communications}\ }\textbf {\bibinfo {volume} {11}},\ \bibinfo {pages} {3860} (\bibinfo {year} {2020})}\BibitemShut {NoStop}%
\bibitem [{\citenamefont {Schulz}\ \emph {et~al.}(2012)\citenamefont {Schulz}, \citenamefont {Ritz}, \citenamefont {Bauer}, \citenamefont {Halder}, \citenamefont {Wagner}, \citenamefont {Franz}, \citenamefont {Pfleiderer}, \citenamefont {Everschor}, \citenamefont {Garst},\ and\ \citenamefont {Rosch}}]{Schulz2012}%
  \BibitemOpen
  \bibfield  {author} {\bibinfo {author} {\bibfnamefont {T.}~\bibnamefont {Schulz}}, \bibinfo {author} {\bibfnamefont {R.}~\bibnamefont {Ritz}}, \bibinfo {author} {\bibfnamefont {A.}~\bibnamefont {Bauer}}, \bibinfo {author} {\bibfnamefont {M.}~\bibnamefont {Halder}}, \bibinfo {author} {\bibfnamefont {M.}~\bibnamefont {Wagner}}, \bibinfo {author} {\bibfnamefont {C.}~\bibnamefont {Franz}}, \bibinfo {author} {\bibfnamefont {C.}~\bibnamefont {Pfleiderer}}, \bibinfo {author} {\bibfnamefont {K.}~\bibnamefont {Everschor}}, \bibinfo {author} {\bibfnamefont {M.}~\bibnamefont {Garst}},\ and\ \bibinfo {author} {\bibfnamefont {A.}~\bibnamefont {Rosch}},\ }\bibfield  {title} {\bibinfo {title} {Emergent electrodynamics of skyrmions in a chiral magnet},\ }\href {https://doi.org/10.1038/nphys2231} {\bibfield  {journal} {\bibinfo  {journal} {Nature Physics}\ }\textbf {\bibinfo {volume} {8}},\ \bibinfo {pages} {301} (\bibinfo {year} {2012})}\BibitemShut {NoStop}%
\bibitem [{\citenamefont {Kanazawa}\ \emph {et~al.}(2011)\citenamefont {Kanazawa}, \citenamefont {Onose}, \citenamefont {Arima}, \citenamefont {Okuyama}, \citenamefont {Ohoyama}, \citenamefont {Wakimoto}, \citenamefont {Kakurai}, \citenamefont {Ishiwata},\ and\ \citenamefont {Tokura}}]{Kanazawa2011}%
  \BibitemOpen
  \bibfield  {author} {\bibinfo {author} {\bibfnamefont {N.}~\bibnamefont {Kanazawa}}, \bibinfo {author} {\bibfnamefont {Y.}~\bibnamefont {Onose}}, \bibinfo {author} {\bibfnamefont {T.}~\bibnamefont {Arima}}, \bibinfo {author} {\bibfnamefont {D.}~\bibnamefont {Okuyama}}, \bibinfo {author} {\bibfnamefont {K.}~\bibnamefont {Ohoyama}}, \bibinfo {author} {\bibfnamefont {S.}~\bibnamefont {Wakimoto}}, \bibinfo {author} {\bibfnamefont {K.}~\bibnamefont {Kakurai}}, \bibinfo {author} {\bibfnamefont {S.}~\bibnamefont {Ishiwata}},\ and\ \bibinfo {author} {\bibfnamefont {Y.}~\bibnamefont {Tokura}},\ }\bibfield  {title} {\bibinfo {title} {Large topological $\mathrm{Hall}$ effect in a short-period helimagnet $\mathrm{MnGe}$},\ }\href {https://doi.org/10.1103/PhysRevLett.106.156603} {\bibfield  {journal} {\bibinfo  {journal} {Physical Review Letters}\ }\textbf {\bibinfo {volume} {106}},\ \bibinfo {pages} {156603} (\bibinfo {year} {2011})}\BibitemShut {NoStop}%
\bibitem [{\citenamefont {Huang}\ and\ \citenamefont {Chien}(2012)}]{Huang2012}%
  \BibitemOpen
  \bibfield  {author} {\bibinfo {author} {\bibfnamefont {S.~X.}\ \bibnamefont {Huang}}\ and\ \bibinfo {author} {\bibfnamefont {C.~L.}\ \bibnamefont {Chien}},\ }\bibfield  {title} {\bibinfo {title} {Extended skyrmion phase in epitaxial $\mathrm{FeGe}$(111) thin films},\ }\href {https://doi.org/10.1103/PhysRevLett.108.267201} {\bibfield  {journal} {\bibinfo  {journal} {Physical Review Letters}\ }\textbf {\bibinfo {volume} {108}},\ \bibinfo {pages} {267201} (\bibinfo {year} {2012})}\BibitemShut {NoStop}%
\bibitem [{\citenamefont {Neubauer}\ \emph {et~al.}(2009)\citenamefont {Neubauer}, \citenamefont {Pfleiderer}, \citenamefont {Binz}, \citenamefont {Rosch}, \citenamefont {Ritz}, \citenamefont {Niklowitz},\ and\ \citenamefont {Böni}}]{Neubauer2009}%
  \BibitemOpen
  \bibfield  {author} {\bibinfo {author} {\bibfnamefont {A.}~\bibnamefont {Neubauer}}, \bibinfo {author} {\bibfnamefont {C.}~\bibnamefont {Pfleiderer}}, \bibinfo {author} {\bibfnamefont {B.}~\bibnamefont {Binz}}, \bibinfo {author} {\bibfnamefont {A.}~\bibnamefont {Rosch}}, \bibinfo {author} {\bibfnamefont {R.}~\bibnamefont {Ritz}}, \bibinfo {author} {\bibfnamefont {P.~G.}\ \bibnamefont {Niklowitz}},\ and\ \bibinfo {author} {\bibfnamefont {P.}~\bibnamefont {Böni}},\ }\bibfield  {title} {\bibinfo {title} {Topological $\mathrm{Hall}$ effect in the $\mathrm{A}$ phase of $\mathrm{MnSi}$},\ }\href {https://doi.org/10.1103/PhysRevLett.102.186602} {\bibfield  {journal} {\bibinfo  {journal} {Physical Review Letters}\ }\textbf {\bibinfo {volume} {102}},\ \bibinfo {pages} {186602} (\bibinfo {year} {2009})}\BibitemShut {NoStop}%
\bibitem [{\citenamefont {Soumyanarayanan}\ \emph {et~al.}(2017)\citenamefont {Soumyanarayanan}, \citenamefont {Raju}, \citenamefont {Oyarce}, \citenamefont {Tan}, \citenamefont {Im}, \citenamefont {Petrović}, \citenamefont {Ho}, \citenamefont {Khoo}, \citenamefont {Tran}, \citenamefont {Gan} \emph {et~al.}}]{Soumyanarayanan2017}%
  \BibitemOpen
  \bibfield  {author} {\bibinfo {author} {\bibfnamefont {A.}~\bibnamefont {Soumyanarayanan}}, \bibinfo {author} {\bibfnamefont {M.}~\bibnamefont {Raju}}, \bibinfo {author} {\bibfnamefont {A.~L.~G.}\ \bibnamefont {Oyarce}}, \bibinfo {author} {\bibfnamefont {A.~K.~C.}\ \bibnamefont {Tan}}, \bibinfo {author} {\bibfnamefont {M.-Y.}\ \bibnamefont {Im}}, \bibinfo {author} {\bibfnamefont {A.~P.}\ \bibnamefont {Petrović}}, \bibinfo {author} {\bibfnamefont {P.}~\bibnamefont {Ho}}, \bibinfo {author} {\bibfnamefont {K.~H.}\ \bibnamefont {Khoo}}, \bibinfo {author} {\bibfnamefont {M.}~\bibnamefont {Tran}}, \bibinfo {author} {\bibfnamefont {C.~K.}\ \bibnamefont {Gan}}, \emph {et~al.},\ }\bibfield  {title} {\bibinfo {title} {Tunable room-temperature magnetic skyrmions in $\mathrm{Ir}/\mathrm{Fe}/\mathrm{Co}/\mathrm{Pt}$ multilayers},\ }\href {https://doi.org/10.1038/nmat4934} {\bibfield  {journal} {\bibinfo  {journal} {Nature Materials}\ }\textbf {\bibinfo {volume} {16}},\ \bibinfo {pages} {898} (\bibinfo {year}
  {2017})}\BibitemShut {NoStop}%
\bibitem [{\citenamefont {He}\ \emph {et~al.}(2018)\citenamefont {He}, \citenamefont {Li}, \citenamefont {Zhu}, \citenamefont {Zhang}, \citenamefont {Peng}, \citenamefont {Li}, \citenamefont {Li}, \citenamefont {Wei}, \citenamefont {Zhao}, \citenamefont {Zhang} \emph {et~al.}}]{He2018}%
  \BibitemOpen
  \bibfield  {author} {\bibinfo {author} {\bibfnamefont {M.}~\bibnamefont {He}}, \bibinfo {author} {\bibfnamefont {G.}~\bibnamefont {Li}}, \bibinfo {author} {\bibfnamefont {Z.}~\bibnamefont {Zhu}}, \bibinfo {author} {\bibfnamefont {Y.}~\bibnamefont {Zhang}}, \bibinfo {author} {\bibfnamefont {L.}~\bibnamefont {Peng}}, \bibinfo {author} {\bibfnamefont {R.}~\bibnamefont {Li}}, \bibinfo {author} {\bibfnamefont {J.}~\bibnamefont {Li}}, \bibinfo {author} {\bibfnamefont {H.}~\bibnamefont {Wei}}, \bibinfo {author} {\bibfnamefont {T.}~\bibnamefont {Zhao}}, \bibinfo {author} {\bibfnamefont {X.-G.}\ \bibnamefont {Zhang}}, \emph {et~al.},\ }\bibfield  {title} {\bibinfo {title} {Evolution of topological skyrmions across the spin reorientation transition in $\mathrm{Pt}/\mathrm{Co}/\mathrm{Ta}$ multilayers},\ }\href {https://doi.org/10.1103/PhysRevB.97.174419} {\bibfield  {journal} {\bibinfo  {journal} {Physical Review B}\ }\textbf {\bibinfo {volume} {97}},\ \bibinfo {pages} {174419} (\bibinfo {year} {2018})}\BibitemShut
  {NoStop}%
\bibitem [{\citenamefont {Mourkas}\ \emph {et~al.}(2021)\citenamefont {Mourkas}, \citenamefont {Markou}, \citenamefont {Swekis},\ and\ \citenamefont {Panagiotopoulos}}]{Mourkas2021}%
  \BibitemOpen
  \bibfield  {author} {\bibinfo {author} {\bibfnamefont {A.}~\bibnamefont {Mourkas}}, \bibinfo {author} {\bibfnamefont {A.}~\bibnamefont {Markou}}, \bibinfo {author} {\bibfnamefont {P.}~\bibnamefont {Swekis}},\ and\ \bibinfo {author} {\bibfnamefont {I.}~\bibnamefont {Panagiotopoulos}},\ }\bibfield  {title} {\bibinfo {title} {Topological hall effect in $\mathrm{Pt}/\mathrm{Co}/\mathrm{W}$ multilayers with different anisotropies},\ }\href {https://doi.org/10.1016/j.jmmm.2021.167937} {\bibfield  {journal} {\bibinfo  {journal} {Journal of Magnetism and Magnetic Materials}\ }\textbf {\bibinfo {volume} {530}},\ \bibinfo {pages} {167937} (\bibinfo {year} {2021})}\BibitemShut {NoStop}%
\bibitem [{\citenamefont {Mohanty}\ \emph {et~al.}(2024)\citenamefont {Mohanty}, \citenamefont {Ojha}, \citenamefont {Sharma},\ and\ \citenamefont {Bedanta}}]{mohanty2024observation}%
  \BibitemOpen
  \bibfield  {author} {\bibinfo {author} {\bibfnamefont {S.}~\bibnamefont {Mohanty}}, \bibinfo {author} {\bibfnamefont {B.}~\bibnamefont {Ojha}}, \bibinfo {author} {\bibfnamefont {M.}~\bibnamefont {Sharma}},\ and\ \bibinfo {author} {\bibfnamefont {S.}~\bibnamefont {Bedanta}},\ }\bibfield  {title} {\bibinfo {title} {Observation of topological $\mathrm{Hall}$ effect and skyrmions in $\mathrm{Pt}/\mathrm{Co}/\mathrm{Ir}/\mathrm{Co}/\mathrm{Pt}$ system},\ }\href {https://dx.doi.org/10.1088/2053-1591/ad3f7a} {\bibfield  {journal} {\bibinfo  {journal} {Materials Research Express}\ }\textbf {\bibinfo {volume} {11}},\ \bibinfo {pages} {046406} (\bibinfo {year} {2024})}\BibitemShut {NoStop}%
\bibitem [{\citenamefont {Jadaun}\ \emph {et~al.}(2020)\citenamefont {Jadaun}, \citenamefont {Register},\ and\ \citenamefont {Banerjee}}]{Jadaun12020}%
  \BibitemOpen
  \bibfield  {author} {\bibinfo {author} {\bibfnamefont {P.}~\bibnamefont {Jadaun}}, \bibinfo {author} {\bibfnamefont {L.~F.}\ \bibnamefont {Register}},\ and\ \bibinfo {author} {\bibfnamefont {S.~K.}\ \bibnamefont {Banerjee}},\ }\bibfield  {title} {\bibinfo {title} {The microscopic origin of $\mathrm{DMI}$ in magnetic bilayers and prediction of giant $\mathrm{DMI}$ in new bilayers},\ }\href {https://doi.org/10.1038/s41524-020-00351-1} {\bibfield  {journal} {\bibinfo  {journal} {NPJ Computational Materials}\ }\textbf {\bibinfo {volume} {6}},\ \bibinfo {pages} {88} (\bibinfo {year} {2020})}\BibitemShut {NoStop}%
\bibitem [{\citenamefont {Fakhredine}\ \emph {et~al.}(2024)\citenamefont {Fakhredine}, \citenamefont {Wawro},\ and\ \citenamefont {Autieri}}]{fakhredine2024huge}%
  \BibitemOpen
  \bibfield  {author} {\bibinfo {author} {\bibfnamefont {A.}~\bibnamefont {Fakhredine}}, \bibinfo {author} {\bibfnamefont {A.}~\bibnamefont {Wawro}},\ and\ \bibinfo {author} {\bibfnamefont {C.}~\bibnamefont {Autieri}},\ }\bibfield  {title} {\bibinfo {title} {Huge $\mathrm{Dzyaloshinskii-Moriya}$ interactions in $\mathrm{Pt}/\mathrm{Co}/\mathrm{Re}$ thin films},\ }\href {https://doi.org/10.1063/5.0177260} {\bibfield  {journal} {\bibinfo  {journal} {Journal of Applied Physics}\ }\textbf {\bibinfo {volume} {135}},\ \bibinfo {pages} {035303} (\bibinfo {year} {2024})}\BibitemShut {NoStop}%
\bibitem [{\citenamefont {Nomura}\ \emph {et~al.}(2022)\citenamefont {Nomura}, \citenamefont {Gao}, \citenamefont {Haku},\ and\ \citenamefont {Ando}}]{Nomura2022}%
  \BibitemOpen
  \bibfield  {author} {\bibinfo {author} {\bibfnamefont {A.}~\bibnamefont {Nomura}}, \bibinfo {author} {\bibfnamefont {T.}~\bibnamefont {Gao}}, \bibinfo {author} {\bibfnamefont {S.}~\bibnamefont {Haku}},\ and\ \bibinfo {author} {\bibfnamefont {K.}~\bibnamefont {Ando}},\ }\bibfield  {title} {\bibinfo {title} {Additive $\mathrm{Dzyaloshinskii-Moriya}$ interaction in $\mathrm{Pt}/\mathrm{Co}/\mathrm{Re}$ films},\ }\href {https://doi.org/10.1063/5.0077683} {\bibfield  {journal} {\bibinfo  {journal} {AIP Advances}\ }\textbf {\bibinfo {volume} {12}},\ \bibinfo {pages} {015215} (\bibinfo {year} {2022})}\BibitemShut {NoStop}%
\bibitem [{\citenamefont {Mallick}\ \emph {et~al.}(2022)\citenamefont {Mallick}, \citenamefont {Panigrahy}, \citenamefont {Pradhan},\ and\ \citenamefont {Rohart}}]{Mallick2022}%
  \BibitemOpen
  \bibfield  {author} {\bibinfo {author} {\bibfnamefont {S.}~\bibnamefont {Mallick}}, \bibinfo {author} {\bibfnamefont {S.}~\bibnamefont {Panigrahy}}, \bibinfo {author} {\bibfnamefont {G.}~\bibnamefont {Pradhan}},\ and\ \bibinfo {author} {\bibfnamefont {S.}~\bibnamefont {Rohart}},\ }\bibfield  {title} {\bibinfo {title} {Current-induced nucleation and motion of skyrmions in zero magnetic field},\ }\href {https://doi.org/10.1103/PhysRevApplied.18.064072} {\bibfield  {journal} {\bibinfo  {journal} {Physical Review Applied}\ }\textbf {\bibinfo {volume} {18}},\ \bibinfo {pages} {064072} (\bibinfo {year} {2022})}\BibitemShut {NoStop}%
\bibitem [{\citenamefont {Behera}\ \emph {et~al.}(2018)\citenamefont {Behera}, \citenamefont {Mishra}, \citenamefont {Mallick}, \citenamefont {Singh},\ and\ \citenamefont {Bedanta}}]{Behera2018}%
  \BibitemOpen
  \bibfield  {author} {\bibinfo {author} {\bibfnamefont {A.~K.}\ \bibnamefont {Behera}}, \bibinfo {author} {\bibfnamefont {S.~S.}\ \bibnamefont {Mishra}}, \bibinfo {author} {\bibfnamefont {S.}~\bibnamefont {Mallick}}, \bibinfo {author} {\bibfnamefont {B.~B.}\ \bibnamefont {Singh}},\ and\ \bibinfo {author} {\bibfnamefont {S.}~\bibnamefont {Bedanta}},\ }\bibfield  {title} {\bibinfo {title} {Size and shape of skyrmions for variable $\mathrm{Dzyaloshinskii-Moriya}$ interaction and uniaxial anisotropy},\ }\href {https://dx.doi.org/10.1088/1361-6463/aac9a7} {\bibfield  {journal} {\bibinfo  {journal} {Journal of Physics D: Applied Physics}\ }\textbf {\bibinfo {volume} {51}},\ \bibinfo {pages} {285001} (\bibinfo {year} {2018})}\BibitemShut {NoStop}%
\bibitem [{\citenamefont {Thiaville}\ \emph {et~al.}(2012)\citenamefont {Thiaville}, \citenamefont {Rohart}, \citenamefont {Émilie Jué}, \citenamefont {Cros},\ and\ \citenamefont {Fert}}]{Thiaville2012}%
  \BibitemOpen
  \bibfield  {author} {\bibinfo {author} {\bibfnamefont {A.}~\bibnamefont {Thiaville}}, \bibinfo {author} {\bibfnamefont {S.}~\bibnamefont {Rohart}}, \bibinfo {author} {\bibnamefont {Émilie Jué}}, \bibinfo {author} {\bibfnamefont {V.}~\bibnamefont {Cros}},\ and\ \bibinfo {author} {\bibfnamefont {A.}~\bibnamefont {Fert}},\ }\bibfield  {title} {\bibinfo {title} {Dynamics of $\mathrm{Dzyaloshinskii}$ domain walls in ultrathin magnetic films},\ }\href {https://dx.doi.org/10.1209/0295-5075/100/57002} {\bibfield  {journal} {\bibinfo  {journal} {EPL (Europhysics Letters)}\ }\textbf {\bibinfo {volume} {100}},\ \bibinfo {pages} {57002} (\bibinfo {year} {2012})}\BibitemShut {NoStop}%
\bibitem [{\citenamefont {Rohart}\ and\ \citenamefont {Thiaville}(2013)}]{Rohart2013}%
  \BibitemOpen
  \bibfield  {author} {\bibinfo {author} {\bibfnamefont {S.}~\bibnamefont {Rohart}}\ and\ \bibinfo {author} {\bibfnamefont {A.}~\bibnamefont {Thiaville}},\ }\bibfield  {title} {\bibinfo {title} {Skyrmion confinement in ultrathin film nanostructures in the presence of $\mathrm{Dzyaloshinskii-Moriya}$ interaction},\ }\href {https://doi.org/10.1103/PhysRevB.88.184422} {\bibfield  {journal} {\bibinfo  {journal} {Physical Review B}\ }\textbf {\bibinfo {volume} {88}},\ \bibinfo {pages} {184422} (\bibinfo {year} {2013})}\BibitemShut {NoStop}%
\bibitem [{\citenamefont {Paul}\ \emph {et~al.}(2020)\citenamefont {Paul}, \citenamefont {Haldar}, \citenamefont {von Malottki},\ and\ \citenamefont {Heinze}}]{paul2020role}%
  \BibitemOpen
  \bibfield  {author} {\bibinfo {author} {\bibfnamefont {S.}~\bibnamefont {Paul}}, \bibinfo {author} {\bibfnamefont {S.}~\bibnamefont {Haldar}}, \bibinfo {author} {\bibfnamefont {S.}~\bibnamefont {von Malottki}},\ and\ \bibinfo {author} {\bibfnamefont {S.}~\bibnamefont {Heinze}},\ }\bibfield  {title} {\bibinfo {title} {Role of higher-order exchange interactions for skyrmion stability},\ }\href {https://doi.org/10.1038/s41467-020-18473-x} {\bibfield  {journal} {\bibinfo  {journal} {Nature Communications}\ }\textbf {\bibinfo {volume} {11}},\ \bibinfo {pages} {4756} (\bibinfo {year} {2020})}\BibitemShut {NoStop}%
\bibitem [{\citenamefont {Jefremovas}\ \emph {et~al.}(2025)\citenamefont {Jefremovas}, \citenamefont {Leutner}, \citenamefont {Fischer}, \citenamefont {Marqu{\'e}s-March{\'a}n}, \citenamefont {Winkler}, \citenamefont {Asenjo}, \citenamefont {Sinova}, \citenamefont {Fr{\"o}mter},\ and\ \citenamefont {Kl{\"a}ui}}]{jefremovas2024role}%
  \BibitemOpen
  \bibfield  {author} {\bibinfo {author} {\bibfnamefont {E.~M.}\ \bibnamefont {Jefremovas}}, \bibinfo {author} {\bibfnamefont {K.}~\bibnamefont {Leutner}}, \bibinfo {author} {\bibfnamefont {M.~G.}\ \bibnamefont {Fischer}}, \bibinfo {author} {\bibfnamefont {J.}~\bibnamefont {Marqu{\'e}s-March{\'a}n}}, \bibinfo {author} {\bibfnamefont {T.~B.}\ \bibnamefont {Winkler}}, \bibinfo {author} {\bibfnamefont {A.}~\bibnamefont {Asenjo}}, \bibinfo {author} {\bibfnamefont {J.}~\bibnamefont {Sinova}}, \bibinfo {author} {\bibfnamefont {R.}~\bibnamefont {Fr{\"o}mter}},\ and\ \bibinfo {author} {\bibfnamefont {M.}~\bibnamefont {Kl{\"a}ui}},\ }\bibfield  {title} {\bibinfo {title} {The role of magnetic dipolar interactions in skyrmion lattices},\ }\href {https://doi.org/10.1016/j.newton.2025.100036} {\bibfield  {journal} {\bibinfo  {journal} {Newton}\ }\textbf {\bibinfo {volume} {1}},\ \bibinfo {pages} {100036} (\bibinfo {year} {2025})}\BibitemShut {NoStop}%
\bibitem [{\citenamefont {Vidal-Silva}\ \emph {et~al.}(2017)\citenamefont {Vidal-Silva}, \citenamefont {Riveros},\ and\ \citenamefont {Escrig}}]{vidal2017stability}%
  \BibitemOpen
  \bibfield  {author} {\bibinfo {author} {\bibfnamefont {N.}~\bibnamefont {Vidal-Silva}}, \bibinfo {author} {\bibfnamefont {A.}~\bibnamefont {Riveros}},\ and\ \bibinfo {author} {\bibfnamefont {J.}~\bibnamefont {Escrig}},\ }\bibfield  {title} {\bibinfo {title} {Stability of neel skyrmions in ultra-thin nanodots considering $\mathrm{Dzyaloshinskii-Moriya}$ and dipolar interactions},\ }\href {https://doi.org/10.1016/j.jmmm.2017.07.049} {\bibfield  {journal} {\bibinfo  {journal} {Journal of Magnetism and Magnetic Materials}\ }\textbf {\bibinfo {volume} {443}},\ \bibinfo {pages} {116} (\bibinfo {year} {2017})}\BibitemShut {NoStop}%
\bibitem [{\citenamefont {Belmeguenai}\ \emph {et~al.}(2015)\citenamefont {Belmeguenai}, \citenamefont {Adam}, \citenamefont {Roussigné}, \citenamefont {Eimer}, \citenamefont {Devolder}, \citenamefont {Kim}, \citenamefont {Cherif}, \citenamefont {Stashkevich},\ and\ \citenamefont {Thiaville}}]{Belmeguenai2015}%
  \BibitemOpen
  \bibfield  {author} {\bibinfo {author} {\bibfnamefont {M.}~\bibnamefont {Belmeguenai}}, \bibinfo {author} {\bibfnamefont {J.-P.}\ \bibnamefont {Adam}}, \bibinfo {author} {\bibfnamefont {Y.}~\bibnamefont {Roussigné}}, \bibinfo {author} {\bibfnamefont {S.}~\bibnamefont {Eimer}}, \bibinfo {author} {\bibfnamefont {T.}~\bibnamefont {Devolder}}, \bibinfo {author} {\bibfnamefont {J.-V.}\ \bibnamefont {Kim}}, \bibinfo {author} {\bibfnamefont {S.~M.}\ \bibnamefont {Cherif}}, \bibinfo {author} {\bibfnamefont {A.}~\bibnamefont {Stashkevich}},\ and\ \bibinfo {author} {\bibfnamefont {A.}~\bibnamefont {Thiaville}},\ }\bibfield  {title} {\bibinfo {title} {Interfacial $\mathrm{Dzyaloshinskii-Moriya}$ interaction in perpendicularly magnetized $\mathrm{Pt}/\mathrm{Co}/\mathrm{AlO_{x}}$ ultrathin films measured by brillouin light spectroscopy},\ }\href {https://doi.org/10.1103/PhysRevB.91.180405} {\bibfield  {journal} {\bibinfo  {journal} {Physical Review B}\ }\textbf {\bibinfo {volume} {91}},\ \bibinfo {pages} {180405}
  (\bibinfo {year} {2015})}\BibitemShut {NoStop}%
\bibitem [{\citenamefont {Shahbazi}\ \emph {et~al.}(2019)\citenamefont {Shahbazi}, \citenamefont {Kim}, \citenamefont {Nembach}, \citenamefont {Shaw}, \citenamefont {Bischof}, \citenamefont {Rossell}, \citenamefont {Jeudy}, \citenamefont {Moore},\ and\ \citenamefont {Marrows}}]{Shahbazi2019}%
  \BibitemOpen
  \bibfield  {author} {\bibinfo {author} {\bibfnamefont {K.}~\bibnamefont {Shahbazi}}, \bibinfo {author} {\bibfnamefont {J.-V.}\ \bibnamefont {Kim}}, \bibinfo {author} {\bibfnamefont {H.~T.}\ \bibnamefont {Nembach}}, \bibinfo {author} {\bibfnamefont {J.~M.}\ \bibnamefont {Shaw}}, \bibinfo {author} {\bibfnamefont {A.}~\bibnamefont {Bischof}}, \bibinfo {author} {\bibfnamefont {M.~D.}\ \bibnamefont {Rossell}}, \bibinfo {author} {\bibfnamefont {V.}~\bibnamefont {Jeudy}}, \bibinfo {author} {\bibfnamefont {T.~A.}\ \bibnamefont {Moore}},\ and\ \bibinfo {author} {\bibfnamefont {C.~H.}\ \bibnamefont {Marrows}},\ }\bibfield  {title} {\bibinfo {title} {Domain-wall motion and interfacial $\mathrm{Dzyaloshinskii-Moriya}$ interactions in $\mathrm{Pt}/\mathrm{Co}/\mathrm{Ir}({t}_{Ir})/\mathrm{Ta}$ multilayers},\ }\href {https://doi.org/10.1103/PhysRevB.99.094409} {\bibfield  {journal} {\bibinfo  {journal} {Physical Review B}\ }\textbf {\bibinfo {volume} {99}},\ \bibinfo {pages} {094409} (\bibinfo {year} {2019})}\BibitemShut
  {NoStop}%
\bibitem [{\citenamefont {Kashid}\ \emph {et~al.}(2014)\citenamefont {Kashid}, \citenamefont {Schena}, \citenamefont {Zimmermann}, \citenamefont {Mokrousov}, \citenamefont {Blügel}, \citenamefont {Shah},\ and\ \citenamefont {Salunke}}]{Kashid2014}%
  \BibitemOpen
  \bibfield  {author} {\bibinfo {author} {\bibfnamefont {V.}~\bibnamefont {Kashid}}, \bibinfo {author} {\bibfnamefont {T.}~\bibnamefont {Schena}}, \bibinfo {author} {\bibfnamefont {B.}~\bibnamefont {Zimmermann}}, \bibinfo {author} {\bibfnamefont {Y.}~\bibnamefont {Mokrousov}}, \bibinfo {author} {\bibfnamefont {S.}~\bibnamefont {Blügel}}, \bibinfo {author} {\bibfnamefont {V.}~\bibnamefont {Shah}},\ and\ \bibinfo {author} {\bibfnamefont {H.~G.}\ \bibnamefont {Salunke}},\ }\bibfield  {title} {\bibinfo {title} {Dzyaloshinskii-moriya interaction and chiral magnetism in $3d\ensuremath{-}5d$ zigzag chains: Tight-binding model and ab initio calculations},\ }\href {https://link.aps.org/doi/10.1103/PhysRevB.90.054412} {\bibfield  {journal} {\bibinfo  {journal} {Physical Review B}\ }\textbf {\bibinfo {volume} {90}},\ \bibinfo {pages} {054412} (\bibinfo {year} {2014})}\BibitemShut {NoStop}%
\bibitem [{\citenamefont {Mallick}\ \emph {et~al.}(2018)\citenamefont {Mallick}, \citenamefont {Mishra},\ and\ \citenamefont {Bedanta}}]{mallick2018relaxation}%
  \BibitemOpen
  \bibfield  {author} {\bibinfo {author} {\bibfnamefont {S.}~\bibnamefont {Mallick}}, \bibinfo {author} {\bibfnamefont {S.~S.}\ \bibnamefont {Mishra}},\ and\ \bibinfo {author} {\bibfnamefont {S.}~\bibnamefont {Bedanta}},\ }\bibfield  {title} {\bibinfo {title} {Relaxation dynamics in magnetic antidot lattice arrays of $\mathrm{Co}/\mathrm{Pt}$ with perpendicular anisotropy},\ }\href {https://doi.org/10.1038/s41598-018-29903-8} {\bibfield  {journal} {\bibinfo  {journal} {Scientific Reports}\ }\textbf {\bibinfo {volume} {8}},\ \bibinfo {pages} {11648} (\bibinfo {year} {2018})}\BibitemShut {NoStop}%
\bibitem [{\citenamefont {Kanak}\ \emph {et~al.}(2007)\citenamefont {Kanak}, \citenamefont {Czapkiewicz}, \citenamefont {Stobiecki}, \citenamefont {Kachel}, \citenamefont {Sveklo}, \citenamefont {Maziewski},\ and\ \citenamefont {Van~Dijken}}]{kanak2007influence}%
  \BibitemOpen
  \bibfield  {author} {\bibinfo {author} {\bibfnamefont {J.}~\bibnamefont {Kanak}}, \bibinfo {author} {\bibfnamefont {M.}~\bibnamefont {Czapkiewicz}}, \bibinfo {author} {\bibfnamefont {T.}~\bibnamefont {Stobiecki}}, \bibinfo {author} {\bibfnamefont {M.}~\bibnamefont {Kachel}}, \bibinfo {author} {\bibfnamefont {I.}~\bibnamefont {Sveklo}}, \bibinfo {author} {\bibfnamefont {A.}~\bibnamefont {Maziewski}},\ and\ \bibinfo {author} {\bibfnamefont {S.}~\bibnamefont {Van~Dijken}},\ }\bibfield  {title} {\bibinfo {title} {Influence of buffer layers on the texture and magnetic properties of $\mathrm{Co}/\mathrm{Pt}$ multilayers with perpendicular anisotropy},\ }\href {https://doi.org/10.1002/pssa.200777104} {\bibfield  {journal} {\bibinfo  {journal} {Physica Status Solidi (a)}\ }\textbf {\bibinfo {volume} {204}},\ \bibinfo {pages} {3950} (\bibinfo {year} {2007})}\BibitemShut {NoStop}%
\bibitem [{\citenamefont {Carcia}(1988)}]{Carcia1988}%
  \BibitemOpen
  \bibfield  {author} {\bibinfo {author} {\bibfnamefont {P.~F.}\ \bibnamefont {Carcia}},\ }\bibfield  {title} {\bibinfo {title} {Perpendicular magnetic anisotropy in $\mathrm{Pd}/\mathrm{Co}$ and $\mathrm{Pt}/\mathrm{Co}$ thin-film layered structures},\ }\href {https://doi.org/10.1063/1.340404} {\bibfield  {journal} {\bibinfo  {journal} {Journal of Applied Physics}\ }\textbf {\bibinfo {volume} {63}},\ \bibinfo {pages} {5066} (\bibinfo {year} {1988})}\BibitemShut {NoStop}%
\bibitem [{\citenamefont {Mangin}\ \emph {et~al.}(2006)\citenamefont {Mangin}, \citenamefont {Ravelosona}, \citenamefont {Katine}, \citenamefont {Carey}, \citenamefont {Terris},\ and\ \citenamefont {Fullerton}}]{mangin2006current}%
  \BibitemOpen
  \bibfield  {author} {\bibinfo {author} {\bibfnamefont {S.}~\bibnamefont {Mangin}}, \bibinfo {author} {\bibfnamefont {D.}~\bibnamefont {Ravelosona}}, \bibinfo {author} {\bibfnamefont {J.}~\bibnamefont {Katine}}, \bibinfo {author} {\bibfnamefont {M.}~\bibnamefont {Carey}}, \bibinfo {author} {\bibfnamefont {B.}~\bibnamefont {Terris}},\ and\ \bibinfo {author} {\bibfnamefont {E.~E.}\ \bibnamefont {Fullerton}},\ }\bibfield  {title} {\bibinfo {title} {Current-induced magnetization reversal in nanopillars with perpendicular anisotropy},\ }\href {https://doi.org/10.1038/nmat1595} {\bibfield  {journal} {\bibinfo  {journal} {Nature Materials}\ }\textbf {\bibinfo {volume} {5}},\ \bibinfo {pages} {210} (\bibinfo {year} {2006})}\BibitemShut {NoStop}%
\bibitem [{\citenamefont {Parakkat}\ \emph {et~al.}(2016)\citenamefont {Parakkat}, \citenamefont {Ganesh},\ and\ \citenamefont {Anil~Kumar}}]{parakkat2016tailoring}%
  \BibitemOpen
  \bibfield  {author} {\bibinfo {author} {\bibfnamefont {V.~M.}\ \bibnamefont {Parakkat}}, \bibinfo {author} {\bibfnamefont {K.}~\bibnamefont {Ganesh}},\ and\ \bibinfo {author} {\bibfnamefont {P.}~\bibnamefont {Anil~Kumar}},\ }\bibfield  {title} {\bibinfo {title} {Tailoring curie temperature and magnetic anisotropy in ultrathin $\mathrm{Pt}/\mathrm{Co}/\mathrm{Pt}$ films},\ }\href {https://doi.org/10.1063/1.4944343} {\bibfield  {journal} {\bibinfo  {journal} {AIP Advances}\ }\textbf {\bibinfo {volume} {6}},\ \bibinfo {pages} {056118} (\bibinfo {year} {2016})}\BibitemShut {NoStop}%
\bibitem [{\citenamefont {Yıldırım}\ \emph {et~al.}(2022)\citenamefont {Yıldırım}, \citenamefont {Marioni}, \citenamefont {Falub}, \citenamefont {Rohrmann}, \citenamefont {Jaeger}, \citenamefont {Rechsteiner}, \citenamefont {Schneider},\ and\ \citenamefont {Hug}}]{yildirim2022tuning}%
  \BibitemOpen
  \bibfield  {author} {\bibinfo {author} {\bibfnamefont {O.}~\bibnamefont {Yıldırım}}, \bibinfo {author} {\bibfnamefont {M.~A.}\ \bibnamefont {Marioni}}, \bibinfo {author} {\bibfnamefont {C.~V.}\ \bibnamefont {Falub}}, \bibinfo {author} {\bibfnamefont {H.}~\bibnamefont {Rohrmann}}, \bibinfo {author} {\bibfnamefont {D.}~\bibnamefont {Jaeger}}, \bibinfo {author} {\bibfnamefont {M.}~\bibnamefont {Rechsteiner}}, \bibinfo {author} {\bibfnamefont {D.}~\bibnamefont {Schneider}},\ and\ \bibinfo {author} {\bibfnamefont {H.~J.}\ \bibnamefont {Hug}},\ }\bibfield  {title} {\bibinfo {title} {Tuning the perpendicular magnetic anisotropy in $\mathrm{Co}/\mathrm{Pt}$ multilayers grown by facing target sputtering and conventional sputtering},\ }\href {https://doi.org/10.1016/j.scriptamat.2021.114285} {\bibfield  {journal} {\bibinfo  {journal} {Scripta Materialia}\ }\textbf {\bibinfo {volume} {207}},\ \bibinfo {pages} {114285} (\bibinfo {year} {2022})}\BibitemShut {NoStop}%
\bibitem [{\citenamefont {Vansteenkiste}\ \emph {et~al.}(2014)\citenamefont {Vansteenkiste}, \citenamefont {Leliaert}, \citenamefont {Dvornik}, \citenamefont {Helsen}, \citenamefont {Garcia-Sanchez},\ and\ \citenamefont {Waeyenberge}}]{Vansteenkiste2014}%
  \BibitemOpen
  \bibfield  {author} {\bibinfo {author} {\bibfnamefont {A.}~\bibnamefont {Vansteenkiste}}, \bibinfo {author} {\bibfnamefont {J.}~\bibnamefont {Leliaert}}, \bibinfo {author} {\bibfnamefont {M.}~\bibnamefont {Dvornik}}, \bibinfo {author} {\bibfnamefont {M.}~\bibnamefont {Helsen}}, \bibinfo {author} {\bibfnamefont {F.}~\bibnamefont {Garcia-Sanchez}},\ and\ \bibinfo {author} {\bibfnamefont {B.~V.}\ \bibnamefont {Waeyenberge}},\ }\bibfield  {title} {\bibinfo {title} {The design and verification of $\mathrm{MuMax3}$},\ }\href {https://doi.org/10.1063/1.4899186} {\bibfield  {journal} {\bibinfo  {journal} {AIP Advances}\ }\textbf {\bibinfo {volume} {4}},\ \bibinfo {pages} {107133} (\bibinfo {year} {2014})}\BibitemShut {NoStop}%
\bibitem [{\citenamefont {Pandey}\ \emph {et~al.}(2023)\citenamefont {Pandey}, \citenamefont {Ojha},\ and\ \citenamefont {Bedanta}}]{Pandey2023}%
  \BibitemOpen
  \bibfield  {author} {\bibinfo {author} {\bibfnamefont {E.}~\bibnamefont {Pandey}}, \bibinfo {author} {\bibfnamefont {B.}~\bibnamefont {Ojha}},\ and\ \bibinfo {author} {\bibfnamefont {S.}~\bibnamefont {Bedanta}},\ }\bibfield  {title} {\bibinfo {title} {Emergence of sizeable interfacial $\mathrm{Dzyaloshinskii-Moriya}$ interaction at cobalt/fullerene interface},\ }\href {https://doi.org/10.1103/PhysRevApplied.19.044013} {\bibfield  {journal} {\bibinfo  {journal} {Physical Review Applied}\ }\textbf {\bibinfo {volume} {19}},\ \bibinfo {pages} {044013} (\bibinfo {year} {2023})}\BibitemShut {NoStop}%
\bibitem [{\citenamefont {Nagaosa}\ and\ \citenamefont {Tokura}(2013)}]{Nagaosa2013}%
  \BibitemOpen
  \bibfield  {author} {\bibinfo {author} {\bibfnamefont {N.}~\bibnamefont {Nagaosa}}\ and\ \bibinfo {author} {\bibfnamefont {Y.}~\bibnamefont {Tokura}},\ }\bibfield  {title} {\bibinfo {title} {Topological properties and dynamics of magnetic skyrmions},\ }\href {https://doi.org/10.1038/nnano.2013.243} {\bibfield  {journal} {\bibinfo  {journal} {Nature Nanotechnology}\ }\textbf {\bibinfo {volume} {8}},\ \bibinfo {pages} {899} (\bibinfo {year} {2013})}\BibitemShut {NoStop}%
\bibitem [{\citenamefont {Johnson}\ \emph {et~al.}(1996)\citenamefont {Johnson}, \citenamefont {Bloemen}, \citenamefont {Broeder},\ and\ \citenamefont {Vries}}]{Johnson1996}%
  \BibitemOpen
  \bibfield  {author} {\bibinfo {author} {\bibfnamefont {M.~T.}\ \bibnamefont {Johnson}}, \bibinfo {author} {\bibfnamefont {P.~J.~H.}\ \bibnamefont {Bloemen}}, \bibinfo {author} {\bibfnamefont {F.~J. A.~D.}\ \bibnamefont {Broeder}},\ and\ \bibinfo {author} {\bibfnamefont {J.~J.~D.}\ \bibnamefont {Vries}},\ }\bibfield  {title} {\bibinfo {title} {Magnetic anisotropy in metallic multilayers},\ }\href {https://dx.doi.org/10.1088/0034-4885/59/11/002} {\bibfield  {journal} {\bibinfo  {journal} {Reports on Progress in Physics}\ }\textbf {\bibinfo {volume} {59}},\ \bibinfo {pages} {1409} (\bibinfo {year} {1996})}\BibitemShut {NoStop}%
\bibitem [{\citenamefont {Berges}\ \emph {et~al.}(2022)\citenamefont {Berges}, \citenamefont {Haltz}, \citenamefont {Panigrahy}, \citenamefont {Mallick}, \citenamefont {Weil}, \citenamefont {Rohart}, \citenamefont {Mougin},\ and\ \citenamefont {Sampaio}}]{Berges2022}%
  \BibitemOpen
  \bibfield  {author} {\bibinfo {author} {\bibfnamefont {L.}~\bibnamefont {Berges}}, \bibinfo {author} {\bibfnamefont {E.}~\bibnamefont {Haltz}}, \bibinfo {author} {\bibfnamefont {S.}~\bibnamefont {Panigrahy}}, \bibinfo {author} {\bibfnamefont {S.}~\bibnamefont {Mallick}}, \bibinfo {author} {\bibfnamefont {R.}~\bibnamefont {Weil}}, \bibinfo {author} {\bibfnamefont {S.}~\bibnamefont {Rohart}}, \bibinfo {author} {\bibfnamefont {A.}~\bibnamefont {Mougin}},\ and\ \bibinfo {author} {\bibfnamefont {J.}~\bibnamefont {Sampaio}},\ }\bibfield  {title} {\bibinfo {title} {Size-dependent mobility of skyrmions beyond pinning in ferrimagnetic $\mathrm{GdCo}$ thin films},\ }\href {https://doi.org/10.1103/PhysRevB.106.144408} {\bibfield  {journal} {\bibinfo  {journal} {Physical Review B}\ }\textbf {\bibinfo {volume} {106}},\ \bibinfo {pages} {144408} (\bibinfo {year} {2022})}\BibitemShut {NoStop}%
\bibitem [{\citenamefont {Sivakumar}\ \emph {et~al.}(2020)\citenamefont {Sivakumar}, \citenamefont {G\"obel}, \citenamefont {Lesne}, \citenamefont {Markou}, \citenamefont {Gidugu}, \citenamefont {Taylor}, \citenamefont {Deniz}, \citenamefont {Jena}, \citenamefont {Felser}, \citenamefont {Mertig} \emph {et~al.}}]{sivakumar2020topological}%
  \BibitemOpen
  \bibfield  {author} {\bibinfo {author} {\bibfnamefont {P.~K.}\ \bibnamefont {Sivakumar}}, \bibinfo {author} {\bibfnamefont {B.}~\bibnamefont {G\"obel}}, \bibinfo {author} {\bibfnamefont {E.}~\bibnamefont {Lesne}}, \bibinfo {author} {\bibfnamefont {A.}~\bibnamefont {Markou}}, \bibinfo {author} {\bibfnamefont {J.}~\bibnamefont {Gidugu}}, \bibinfo {author} {\bibfnamefont {J.~M.}\ \bibnamefont {Taylor}}, \bibinfo {author} {\bibfnamefont {H.}~\bibnamefont {Deniz}}, \bibinfo {author} {\bibfnamefont {J.}~\bibnamefont {Jena}}, \bibinfo {author} {\bibfnamefont {C.}~\bibnamefont {Felser}}, \bibinfo {author} {\bibfnamefont {I.}~\bibnamefont {Mertig}}, \emph {et~al.},\ }\bibfield  {title} {\bibinfo {title} {Topological $\mathrm{Hall}$ signatures of two chiral spin textures hosted in a single tetragonal inverse heusler thin film},\ }\href {https://doi.org/10.1021/acsnano.0c05413} {\bibfield  {journal} {\bibinfo  {journal} {ACS Nano}\ }\textbf {\bibinfo {volume} {14}},\ \bibinfo {pages} {13463} (\bibinfo {year}
  {2020})}\BibitemShut {NoStop}%
\bibitem [{\citenamefont {Li}\ \emph {et~al.}(2019)\citenamefont {Li}, \citenamefont {Zhang}, \citenamefont {Zhang}, \citenamefont {Li}, \citenamefont {Yang}, \citenamefont {Deng}, \citenamefont {Gu},\ and\ \citenamefont {Wu}}]{Li2019}%
  \BibitemOpen
  \bibfield  {author} {\bibinfo {author} {\bibfnamefont {Y.}~\bibnamefont {Li}}, \bibinfo {author} {\bibfnamefont {L.}~\bibnamefont {Zhang}}, \bibinfo {author} {\bibfnamefont {Q.}~\bibnamefont {Zhang}}, \bibinfo {author} {\bibfnamefont {C.}~\bibnamefont {Li}}, \bibinfo {author} {\bibfnamefont {T.}~\bibnamefont {Yang}}, \bibinfo {author} {\bibfnamefont {Y.}~\bibnamefont {Deng}}, \bibinfo {author} {\bibfnamefont {L.}~\bibnamefont {Gu}},\ and\ \bibinfo {author} {\bibfnamefont {D.}~\bibnamefont {Wu}},\ }\bibfield  {title} {\bibinfo {title} {Emergent topological $\mathrm{Hall}$ effect in $\mathrm{La_{0.7}Sr_{0.3}MnO_{3}}/\mathrm{SrIrO_{3}}$ heterostructures},\ }\href {https://doi.org/10.1021/acsami.9b05562} {\bibfield  {journal} {\bibinfo  {journal} {ACS Applied Materials and Interfaces}\ }\textbf {\bibinfo {volume} {11}},\ \bibinfo {pages} {21268} (\bibinfo {year} {2019})}\BibitemShut {NoStop}%
\bibitem [{\citenamefont {Raju}\ \emph {et~al.}(2021)\citenamefont {Raju}, \citenamefont {Petrović}, \citenamefont {Yagil}, \citenamefont {Denisov}, \citenamefont {Duong}, \citenamefont {Göbel}, \citenamefont {Şaşoğlu}, \citenamefont {Auslaender}, \citenamefont {Mertig}, \citenamefont {Rozhansky} \emph {et~al.}}]{Raju2021}%
  \BibitemOpen
  \bibfield  {author} {\bibinfo {author} {\bibfnamefont {M.}~\bibnamefont {Raju}}, \bibinfo {author} {\bibfnamefont {A.~P.}\ \bibnamefont {Petrović}}, \bibinfo {author} {\bibfnamefont {A.}~\bibnamefont {Yagil}}, \bibinfo {author} {\bibfnamefont {K.~S.}\ \bibnamefont {Denisov}}, \bibinfo {author} {\bibfnamefont {N.~K.}\ \bibnamefont {Duong}}, \bibinfo {author} {\bibfnamefont {B.}~\bibnamefont {Göbel}}, \bibinfo {author} {\bibfnamefont {E.}~\bibnamefont {Şaşoğlu}}, \bibinfo {author} {\bibfnamefont {O.~M.}\ \bibnamefont {Auslaender}}, \bibinfo {author} {\bibfnamefont {I.}~\bibnamefont {Mertig}}, \bibinfo {author} {\bibfnamefont {I.~V.}\ \bibnamefont {Rozhansky}}, \emph {et~al.},\ }\bibfield  {title} {\bibinfo {title} {Colossal topological $\mathrm{Hall}$ effect at the transition between isolated and lattice-phase interfacial skyrmions},\ }\href {https://doi.org/10.1038/s41467-021-22976-6} {\bibfield  {journal} {\bibinfo  {journal} {Nature Communications}\ }\textbf {\bibinfo {volume} {12}},\ \bibinfo {pages}
  {2758} (\bibinfo {year} {2021})}\BibitemShut {NoStop}%
\bibitem [{\citenamefont {Raju}\ \emph {et~al.}(2019)\citenamefont {Raju}, \citenamefont {Yagil}, \citenamefont {Soumyanarayanan}, \citenamefont {Tan}, \citenamefont {Almoalem}, \citenamefont {Ma}, \citenamefont {Auslaender},\ and\ \citenamefont {Panagopoulos}}]{raju2019evolution}%
  \BibitemOpen
  \bibfield  {author} {\bibinfo {author} {\bibfnamefont {M.}~\bibnamefont {Raju}}, \bibinfo {author} {\bibfnamefont {A.}~\bibnamefont {Yagil}}, \bibinfo {author} {\bibfnamefont {A.}~\bibnamefont {Soumyanarayanan}}, \bibinfo {author} {\bibfnamefont {A.~K.}\ \bibnamefont {Tan}}, \bibinfo {author} {\bibfnamefont {A.}~\bibnamefont {Almoalem}}, \bibinfo {author} {\bibfnamefont {F.}~\bibnamefont {Ma}}, \bibinfo {author} {\bibfnamefont {O.}~\bibnamefont {Auslaender}},\ and\ \bibinfo {author} {\bibfnamefont {C.}~\bibnamefont {Panagopoulos}},\ }\bibfield  {title} {\bibinfo {title} {The evolution of skyrmions in $\mathrm{Ir}/\mathrm{Fe}/\mathrm{Co}/\mathrm{Pt}$ multilayers and their topological $\mathrm{Hall}$ signature},\ }\href {https://doi.org/10.1038/s41467-018-08041-9} {\bibfield  {journal} {\bibinfo  {journal} {Nature Communications}\ }\textbf {\bibinfo {volume} {10}},\ \bibinfo {pages} {696} (\bibinfo {year} {2019})}\BibitemShut {NoStop}%
\bibitem [{\citenamefont {Li}\ \emph {et~al.}(2020)\citenamefont {Li}, \citenamefont {Ding}, \citenamefont {Zhang}, \citenamefont {Kally}, \citenamefont {Pillsbury}, \citenamefont {Heinonen}, \citenamefont {Rimal}, \citenamefont {Bi}, \citenamefont {DeMann}, \citenamefont {Field} \emph {et~al.}}]{Li2020}%
  \BibitemOpen
  \bibfield  {author} {\bibinfo {author} {\bibfnamefont {P.}~\bibnamefont {Li}}, \bibinfo {author} {\bibfnamefont {J.}~\bibnamefont {Ding}}, \bibinfo {author} {\bibfnamefont {S.~S.-L.}\ \bibnamefont {Zhang}}, \bibinfo {author} {\bibfnamefont {J.}~\bibnamefont {Kally}}, \bibinfo {author} {\bibfnamefont {T.}~\bibnamefont {Pillsbury}}, \bibinfo {author} {\bibfnamefont {O.~G.}\ \bibnamefont {Heinonen}}, \bibinfo {author} {\bibfnamefont {G.}~\bibnamefont {Rimal}}, \bibinfo {author} {\bibfnamefont {C.}~\bibnamefont {Bi}}, \bibinfo {author} {\bibfnamefont {A.}~\bibnamefont {DeMann}}, \bibinfo {author} {\bibfnamefont {S.~B.}\ \bibnamefont {Field}}, \emph {et~al.},\ }\bibfield  {title} {\bibinfo {title} {Topological $\mathrm{Hall}$ effect in a topological insulator interfaced with a magnetic insulator},\ }\href {https://doi.org/10.1021/acs.nanolett.0c03195} {\bibfield  {journal} {\bibinfo  {journal} {Nano Letters}\ }\textbf {\bibinfo {volume} {21}},\ \bibinfo {pages} {84} (\bibinfo {year} {2020})}\BibitemShut
  {NoStop}%
\bibitem [{\citenamefont {Meng}\ \emph {et~al.}(2018)\citenamefont {Meng}, \citenamefont {Zhao}, \citenamefont {Liu}, \citenamefont {Liu}, \citenamefont {Wu}, \citenamefont {Li}, \citenamefont {Chen}, \citenamefont {Miao}, \citenamefont {Xu}, \citenamefont {Zhao} \emph {et~al.}}]{Meng2018}%
  \BibitemOpen
  \bibfield  {author} {\bibinfo {author} {\bibfnamefont {K.~K.}\ \bibnamefont {Meng}}, \bibinfo {author} {\bibfnamefont {X.~P.}\ \bibnamefont {Zhao}}, \bibinfo {author} {\bibfnamefont {P.~F.}\ \bibnamefont {Liu}}, \bibinfo {author} {\bibfnamefont {Q.}~\bibnamefont {Liu}}, \bibinfo {author} {\bibfnamefont {Y.}~\bibnamefont {Wu}}, \bibinfo {author} {\bibfnamefont {Z.~P.}\ \bibnamefont {Li}}, \bibinfo {author} {\bibfnamefont {J.~K.}\ \bibnamefont {Chen}}, \bibinfo {author} {\bibfnamefont {J.}~\bibnamefont {Miao}}, \bibinfo {author} {\bibfnamefont {X.~G.}\ \bibnamefont {Xu}}, \bibinfo {author} {\bibfnamefont {J.~H.}\ \bibnamefont {Zhao}}, \emph {et~al.},\ }\bibfield  {title} {\bibinfo {title} {Robust emergence of a topological $\mathrm{Hall}$ effect in $\mathrm{MnGa}$/heavy metal bilayers},\ }\href {https://doi.org/10.1103/PhysRevB.97.060407} {\bibfield  {journal} {\bibinfo  {journal} {Physical Review B}\ }\textbf {\bibinfo {volume} {97}},\ \bibinfo {pages} {060407} (\bibinfo {year} {2018})}\BibitemShut {NoStop}%
\bibitem [{\citenamefont {Budhathoki}\ \emph {et~al.}(2020)\citenamefont {Budhathoki}, \citenamefont {Sapkota}, \citenamefont {Law}, \citenamefont {Ranjit}, \citenamefont {Nepal}, \citenamefont {Hoskins}, \citenamefont {Thind}, \citenamefont {Borisevich}, \citenamefont {Jamer}, \citenamefont {Anderson} \emph {et~al.}}]{Budhathoki2020}%
  \BibitemOpen
  \bibfield  {author} {\bibinfo {author} {\bibfnamefont {S.}~\bibnamefont {Budhathoki}}, \bibinfo {author} {\bibfnamefont {A.}~\bibnamefont {Sapkota}}, \bibinfo {author} {\bibfnamefont {K.~M.}\ \bibnamefont {Law}}, \bibinfo {author} {\bibfnamefont {S.}~\bibnamefont {Ranjit}}, \bibinfo {author} {\bibfnamefont {B.}~\bibnamefont {Nepal}}, \bibinfo {author} {\bibfnamefont {B.~D.}\ \bibnamefont {Hoskins}}, \bibinfo {author} {\bibfnamefont {A.~S.}\ \bibnamefont {Thind}}, \bibinfo {author} {\bibfnamefont {A.~Y.}\ \bibnamefont {Borisevich}}, \bibinfo {author} {\bibfnamefont {M.~E.}\ \bibnamefont {Jamer}}, \bibinfo {author} {\bibfnamefont {T.~J.}\ \bibnamefont {Anderson}}, \emph {et~al.},\ }\bibfield  {title} {\bibinfo {title} {Room-temperature skyrmions in strain-engineered $\mathrm{FeGe}$ thin films},\ }\href {https://doi.org/10.1103/PhysRevB.101.220405} {\bibfield  {journal} {\bibinfo  {journal} {Physical Review B}\ }\textbf {\bibinfo {volume} {101}},\ \bibinfo {pages} {220405} (\bibinfo {year} {2020})}\BibitemShut
  {NoStop}%
\bibitem [{\citenamefont {Ahmed}\ \emph {et~al.}(2019)\citenamefont {Ahmed}, \citenamefont {Lee}, \citenamefont {Bagués}, \citenamefont {McCullian}, \citenamefont {Thabt}, \citenamefont {Perrine}, \citenamefont {Wu}, \citenamefont {Rowland}, \citenamefont {Randeria}, \citenamefont {Hammel} \emph {et~al.}}]{Ahmed2019}%
  \BibitemOpen
  \bibfield  {author} {\bibinfo {author} {\bibfnamefont {A.~S.}\ \bibnamefont {Ahmed}}, \bibinfo {author} {\bibfnamefont {A.~J.}\ \bibnamefont {Lee}}, \bibinfo {author} {\bibfnamefont {N.}~\bibnamefont {Bagués}}, \bibinfo {author} {\bibfnamefont {B.~A.}\ \bibnamefont {McCullian}}, \bibinfo {author} {\bibfnamefont {A.~M.~A.}\ \bibnamefont {Thabt}}, \bibinfo {author} {\bibfnamefont {A.}~\bibnamefont {Perrine}}, \bibinfo {author} {\bibfnamefont {P.-K.}\ \bibnamefont {Wu}}, \bibinfo {author} {\bibfnamefont {J.~R.}\ \bibnamefont {Rowland}}, \bibinfo {author} {\bibfnamefont {M.}~\bibnamefont {Randeria}}, \bibinfo {author} {\bibfnamefont {P.~C.}\ \bibnamefont {Hammel}}, \emph {et~al.},\ }\bibfield  {title} {\bibinfo {title} {Spin-$\mathrm{Hall}$ topological $\mathrm{Hall}$ effect in highly tunable $\mathrm{Pt}$/ferrimagnetic-insulator bilayers},\ }\href {https://doi.org/10.1021/acs.nanolett.9b02265} {\bibfield  {journal} {\bibinfo  {journal} {Nano Letters}\ }\textbf {\bibinfo {volume} {19}},\ \bibinfo {pages}
  {5683} (\bibinfo {year} {2019})}\BibitemShut {NoStop}%
\bibitem [{\citenamefont {Akhtar}\ \emph {et~al.}(2019)\citenamefont {Akhtar}, \citenamefont {Hrabec}, \citenamefont {Chouaieb}, \citenamefont {Haykal}, \citenamefont {Gross}, \citenamefont {Belmeguenai}, \citenamefont {Gabor}, \citenamefont {Shields}, \citenamefont {Maletinsky}, \citenamefont {Thiaville} \emph {et~al.}}]{Akhtar2019}%
  \BibitemOpen
  \bibfield  {author} {\bibinfo {author} {\bibfnamefont {W.}~\bibnamefont {Akhtar}}, \bibinfo {author} {\bibfnamefont {A.}~\bibnamefont {Hrabec}}, \bibinfo {author} {\bibfnamefont {S.}~\bibnamefont {Chouaieb}}, \bibinfo {author} {\bibfnamefont {A.}~\bibnamefont {Haykal}}, \bibinfo {author} {\bibfnamefont {I.}~\bibnamefont {Gross}}, \bibinfo {author} {\bibfnamefont {M.}~\bibnamefont {Belmeguenai}}, \bibinfo {author} {\bibfnamefont {M.~S.}\ \bibnamefont {Gabor}}, \bibinfo {author} {\bibfnamefont {B.}~\bibnamefont {Shields}}, \bibinfo {author} {\bibfnamefont {P.}~\bibnamefont {Maletinsky}}, \bibinfo {author} {\bibfnamefont {A.}~\bibnamefont {Thiaville}}, \emph {et~al.},\ }\bibfield  {title} {\bibinfo {title} {Current-induced nucleation and dynamics of skyrmions in a $\mathrm{Co}$-based heusler alloy},\ }\href {https://doi.org/10.1103/PhysRevApplied.11.034066} {\bibfield  {journal} {\bibinfo  {journal} {Physical Review Applied}\ }\textbf {\bibinfo {volume} {11}},\ \bibinfo {pages} {034066} (\bibinfo {year}
  {2019})}\BibitemShut {NoStop}%
\bibitem [{\citenamefont {Ajejas}\ \emph {et~al.}(2023)\citenamefont {Ajejas}, \citenamefont {Sassi}, \citenamefont {Legrand}, \citenamefont {Srivastava}, \citenamefont {Collin}, \citenamefont {Vecchiola}, \citenamefont {Bouzehouane}, \citenamefont {Reyren},\ and\ \citenamefont {Cros}}]{ajejas2023densely}%
  \BibitemOpen
  \bibfield  {author} {\bibinfo {author} {\bibfnamefont {F.}~\bibnamefont {Ajejas}}, \bibinfo {author} {\bibfnamefont {Y.}~\bibnamefont {Sassi}}, \bibinfo {author} {\bibfnamefont {W.}~\bibnamefont {Legrand}}, \bibinfo {author} {\bibfnamefont {T.}~\bibnamefont {Srivastava}}, \bibinfo {author} {\bibfnamefont {S.}~\bibnamefont {Collin}}, \bibinfo {author} {\bibfnamefont {A.}~\bibnamefont {Vecchiola}}, \bibinfo {author} {\bibfnamefont {K.}~\bibnamefont {Bouzehouane}}, \bibinfo {author} {\bibfnamefont {N.}~\bibnamefont {Reyren}},\ and\ \bibinfo {author} {\bibfnamefont {V.}~\bibnamefont {Cros}},\ }\bibfield  {title} {\bibinfo {title} {Densely packed skyrmions stabilized at zero magnetic field by indirect exchange coupling in multilayers},\ }\href {https://doi.org/10.1063/5.0139283} {\bibfield  {journal} {\bibinfo  {journal} {APL Materials}\ }\textbf {\bibinfo {volume} {11}},\ \bibinfo {pages} {061108} (\bibinfo {year} {2023})}\BibitemShut {NoStop}%
\bibitem [{\citenamefont {Sampaio}\ \emph {et~al.}(2013)\citenamefont {Sampaio}, \citenamefont {Cros}, \citenamefont {Rohart}, \citenamefont {Thiaville},\ and\ \citenamefont {Fert}}]{Sampaio2013}%
  \BibitemOpen
  \bibfield  {author} {\bibinfo {author} {\bibfnamefont {J.}~\bibnamefont {Sampaio}}, \bibinfo {author} {\bibfnamefont {V.}~\bibnamefont {Cros}}, \bibinfo {author} {\bibfnamefont {S.}~\bibnamefont {Rohart}}, \bibinfo {author} {\bibfnamefont {A.}~\bibnamefont {Thiaville}},\ and\ \bibinfo {author} {\bibfnamefont {A.}~\bibnamefont {Fert}},\ }\bibfield  {title} {\bibinfo {title} {Nucleation, stability and current-induced motion of isolated magnetic skyrmions in nanostructures},\ }\href {https://doi.org/10.1038/nnano.2013.210} {\bibfield  {journal} {\bibinfo  {journal} {Nature Nanotechnology}\ }\textbf {\bibinfo {volume} {8}},\ \bibinfo {pages} {839} (\bibinfo {year} {2013})}\BibitemShut {NoStop}%
\bibitem [{\citenamefont {Chui}\ \emph {et~al.}(2015)\citenamefont {Chui}, \citenamefont {Ma},\ and\ \citenamefont {Zhou}}]{chui2015geometrical}%
  \BibitemOpen
  \bibfield  {author} {\bibinfo {author} {\bibfnamefont {C.}~\bibnamefont {Chui}}, \bibinfo {author} {\bibfnamefont {F.}~\bibnamefont {Ma}},\ and\ \bibinfo {author} {\bibfnamefont {Y.}~\bibnamefont {Zhou}},\ }\bibfield  {title} {\bibinfo {title} {Geometrical and physical conditions for skyrmion stability in a nanowire},\ }\href {https://doi.org/10.1063/1.4919320} {\bibfield  {journal} {\bibinfo  {journal} {AIP Advances}\ }\textbf {\bibinfo {volume} {5}},\ \bibinfo {pages} {047141} (\bibinfo {year} {2015})}\BibitemShut {NoStop}%
\bibitem [{\citenamefont {Davydenko}\ \emph {et~al.}(2019)\citenamefont {Davydenko}, \citenamefont {Kozlov}, \citenamefont {Kolesnikov}, \citenamefont {Stebliy}, \citenamefont {Suslin}, \citenamefont {Vekovshinin}, \citenamefont {Sadovnikov},\ and\ \citenamefont {Nikitov}}]{Davydenko2019}%
  \BibitemOpen
  \bibfield  {author} {\bibinfo {author} {\bibfnamefont {A.~V.}\ \bibnamefont {Davydenko}}, \bibinfo {author} {\bibfnamefont {A.~G.}\ \bibnamefont {Kozlov}}, \bibinfo {author} {\bibfnamefont {A.~G.}\ \bibnamefont {Kolesnikov}}, \bibinfo {author} {\bibfnamefont {M.~E.}\ \bibnamefont {Stebliy}}, \bibinfo {author} {\bibfnamefont {G.~S.}\ \bibnamefont {Suslin}}, \bibinfo {author} {\bibfnamefont {Y.~E.}\ \bibnamefont {Vekovshinin}}, \bibinfo {author} {\bibfnamefont {A.~V.}\ \bibnamefont {Sadovnikov}},\ and\ \bibinfo {author} {\bibfnamefont {S.~A.}\ \bibnamefont {Nikitov}},\ }\bibfield  {title} {\bibinfo {title} {$\mathrm{Dzyaloshinskii-Moriya}$ interaction in symmetric epitaxial ${[\mathrm{Co}/\mathrm{Pd}(111)]}_{N}$ superlattices with different numbers of $\mathrm{Co}/\mathrm{Pd}$ bilayers},\ }\href {https://doi.org/10.1103/PhysRevB.99.014433} {\bibfield  {journal} {\bibinfo  {journal} {Physical Review B}\ }\textbf {\bibinfo {volume} {99}},\ \bibinfo {pages} {014433} (\bibinfo {year} {2019})}\BibitemShut
  {NoStop}%
\bibitem [{\citenamefont {Ding}\ \emph {et~al.}(2005)\citenamefont {Ding}, \citenamefont {Schmid}, \citenamefont {Li}, \citenamefont {Guslienko},\ and\ \citenamefont {Bader}}]{ding2005magnetic}%
  \BibitemOpen
  \bibfield  {author} {\bibinfo {author} {\bibfnamefont {H.}~\bibnamefont {Ding}}, \bibinfo {author} {\bibfnamefont {A.}~\bibnamefont {Schmid}}, \bibinfo {author} {\bibfnamefont {D.}~\bibnamefont {Li}}, \bibinfo {author} {\bibfnamefont {K.~Y.}\ \bibnamefont {Guslienko}},\ and\ \bibinfo {author} {\bibfnamefont {S.}~\bibnamefont {Bader}},\ }\bibfield  {title} {\bibinfo {title} {Magnetic bistability of $\mathrm{Co}$ nanodots},\ }\href {https://doi.org/10.1103/PhysRevLett.94.157202} {\bibfield  {journal} {\bibinfo  {journal} {Physical Review Letters}\ }\textbf {\bibinfo {volume} {94}},\ \bibinfo {pages} {157202} (\bibinfo {year} {2005})}\BibitemShut {NoStop}%
\bibitem [{\citenamefont {Vernon}\ \emph {et~al.}(1984)\citenamefont {Vernon}, \citenamefont {Lindsay},\ and\ \citenamefont {Stearns}}]{vernon1984brillouin}%
  \BibitemOpen
  \bibfield  {author} {\bibinfo {author} {\bibfnamefont {S.}~\bibnamefont {Vernon}}, \bibinfo {author} {\bibfnamefont {S.}~\bibnamefont {Lindsay}},\ and\ \bibinfo {author} {\bibfnamefont {M.}~\bibnamefont {Stearns}},\ }\bibfield  {title} {\bibinfo {title} {Brillouin scattering from thermal magnons in a thin $\mathrm{Co}$ film},\ }\href {https://doi.org/10.1103/PhysRevB.29.4439} {\bibfield  {journal} {\bibinfo  {journal} {Physical Review B}\ }\textbf {\bibinfo {volume} {29}},\ \bibinfo {pages} {4439} (\bibinfo {year} {1984})}\BibitemShut {NoStop}%
\bibitem [{\citenamefont {B{\"o}ttcher}\ \emph {et~al.}(2023)\citenamefont {B{\"o}ttcher}, \citenamefont {Suraj}, \citenamefont {Chen}, \citenamefont {Sinha}, \citenamefont {Tan}, \citenamefont {Tan}, \citenamefont {Laskowski}, \citenamefont {Hillebrands}, \citenamefont {Kostylev}, \citenamefont {Khoo} \emph {et~al.}}]{bottcher2023quantifying}%
  \BibitemOpen
  \bibfield  {author} {\bibinfo {author} {\bibfnamefont {T.}~\bibnamefont {B{\"o}ttcher}}, \bibinfo {author} {\bibfnamefont {T.~S.}\ \bibnamefont {Suraj}}, \bibinfo {author} {\bibfnamefont {X.}~\bibnamefont {Chen}}, \bibinfo {author} {\bibfnamefont {B.}~\bibnamefont {Sinha}}, \bibinfo {author} {\bibfnamefont {H.~R.}\ \bibnamefont {Tan}}, \bibinfo {author} {\bibfnamefont {H.~K.}\ \bibnamefont {Tan}}, \bibinfo {author} {\bibfnamefont {R.}~\bibnamefont {Laskowski}}, \bibinfo {author} {\bibfnamefont {B.}~\bibnamefont {Hillebrands}}, \bibinfo {author} {\bibfnamefont {M.}~\bibnamefont {Kostylev}}, \bibinfo {author} {\bibfnamefont {K.~H.}\ \bibnamefont {Khoo}}, \emph {et~al.},\ }\bibfield  {title} {\bibinfo {title} {Quantifying symmetric exchange in ultrathin ferromagnetic films with chirality},\ }\href {https://doi.org/10.1103/PhysRevB.107.094405} {\bibfield  {journal} {\bibinfo  {journal} {Physical Review B}\ }\textbf {\bibinfo {volume} {107}},\ \bibinfo {pages} {094405} (\bibinfo {year} {2023})}\BibitemShut
  {NoStop}%
\bibitem [{\citenamefont {Tomasello}\ \emph {et~al.}(2014{\natexlab{b}})\citenamefont {Tomasello}, \citenamefont {Martinez}, \citenamefont {Zivieri}, \citenamefont {Torres}, \citenamefont {Carpentieri},\ and\ \citenamefont {Finocchio}}]{tomasello2014strategy}%
  \BibitemOpen
  \bibfield  {author} {\bibinfo {author} {\bibfnamefont {R.}~\bibnamefont {Tomasello}}, \bibinfo {author} {\bibfnamefont {E.}~\bibnamefont {Martinez}}, \bibinfo {author} {\bibfnamefont {R.}~\bibnamefont {Zivieri}}, \bibinfo {author} {\bibfnamefont {L.}~\bibnamefont {Torres}}, \bibinfo {author} {\bibfnamefont {M.}~\bibnamefont {Carpentieri}},\ and\ \bibinfo {author} {\bibfnamefont {G.}~\bibnamefont {Finocchio}},\ }\bibfield  {title} {\bibinfo {title} {A strategy for the design of skyrmion racetrack memories},\ }\href {https://doi.org/10.1038/srep06784} {\bibfield  {journal} {\bibinfo  {journal} {Scientific Reports}\ }\textbf {\bibinfo {volume} {4}},\ \bibinfo {pages} {6784} (\bibinfo {year} {2014}{\natexlab{b}})}\BibitemShut {NoStop}%
\bibitem [{\citenamefont {Luo}\ and\ \citenamefont {You}(2021{\natexlab{b}})}]{luo2021skyrmion}%
  \BibitemOpen
  \bibfield  {author} {\bibinfo {author} {\bibfnamefont {S.}~\bibnamefont {Luo}}\ and\ \bibinfo {author} {\bibfnamefont {L.}~\bibnamefont {You}},\ }\bibfield  {title} {\bibinfo {title} {Skyrmion devices for memory and logic applications},\ }\href {https://doi.org/10.1063/5.0042917} {\bibfield  {journal} {\bibinfo  {journal} {APL Materials}\ }\textbf {\bibinfo {volume} {9}},\ \bibinfo {pages} {050901} (\bibinfo {year} {2021}{\natexlab{b}})}\BibitemShut {NoStop}%
\bibitem [{\citenamefont {Zhang}\ \emph {et~al.}(2015)\citenamefont {Zhang}, \citenamefont {Ezawa},\ and\ \citenamefont {Zhou}}]{zhang2015magnetic}%
  \BibitemOpen
  \bibfield  {author} {\bibinfo {author} {\bibfnamefont {X.}~\bibnamefont {Zhang}}, \bibinfo {author} {\bibfnamefont {M.}~\bibnamefont {Ezawa}},\ and\ \bibinfo {author} {\bibfnamefont {Y.}~\bibnamefont {Zhou}},\ }\bibfield  {title} {\bibinfo {title} {Magnetic skyrmion logic gates: conversion, duplication and merging of skyrmions},\ }\href {https://doi.org/10.1038/srep09400} {\bibfield  {journal} {\bibinfo  {journal} {Scientific Reports}\ }\textbf {\bibinfo {volume} {5}},\ \bibinfo {pages} {9400} (\bibinfo {year} {2015})}\BibitemShut {NoStop}%
\bibitem [{\citenamefont {Raab}\ \emph {et~al.}(2022)\citenamefont {Raab}, \citenamefont {Brems}, \citenamefont {Beneke}, \citenamefont {Dohi}, \citenamefont {Roth{\"o}rl}, \citenamefont {Kammerbauer}, \citenamefont {Mentink},\ and\ \citenamefont {Kl{\"a}ui}}]{raab2022brownian}%
  \BibitemOpen
  \bibfield  {author} {\bibinfo {author} {\bibfnamefont {K.}~\bibnamefont {Raab}}, \bibinfo {author} {\bibfnamefont {M.~A.}\ \bibnamefont {Brems}}, \bibinfo {author} {\bibfnamefont {G.}~\bibnamefont {Beneke}}, \bibinfo {author} {\bibfnamefont {T.}~\bibnamefont {Dohi}}, \bibinfo {author} {\bibfnamefont {J.}~\bibnamefont {Roth{\"o}rl}}, \bibinfo {author} {\bibfnamefont {F.}~\bibnamefont {Kammerbauer}}, \bibinfo {author} {\bibfnamefont {J.~H.}\ \bibnamefont {Mentink}},\ and\ \bibinfo {author} {\bibfnamefont {M.}~\bibnamefont {Kl{\"a}ui}},\ }\bibfield  {title} {\bibinfo {title} {Brownian reservoir computing realized using geometrically confined skyrmion dynamics},\ }\href {https://doi.org/10.1038/s41467-022-34309-2} {\bibfield  {journal} {\bibinfo  {journal} {Nature Communications}\ }\textbf {\bibinfo {volume} {13}},\ \bibinfo {pages} {6982} (\bibinfo {year} {2022})}\BibitemShut {NoStop}%
\end{thebibliography}%


\begin{thebibliography}{5}%
\makeatletter
\providecommand \@ifxundefined [1]{%
 \@ifx{#1\undefined}
}%
\providecommand \@ifnum [1]{%
 \ifnum #1\expandafter \@firstoftwo
 \else \expandafter \@secondoftwo
 \fi
}%
\providecommand \@ifx [1]{%
 \ifx #1\expandafter \@firstoftwo
 \else \expandafter \@secondoftwo
 \fi
}%
\providecommand \natexlab [1]{#1}%
\providecommand \enquote  [1]{``#1''}%
\providecommand \bibnamefont  [1]{#1}%
\providecommand \bibfnamefont [1]{#1}%
\providecommand \citenamefont [1]{#1}%
\providecommand \href@noop [0]{\@secondoftwo}%
\providecommand \href [0]{\begingroup \@sanitize@url \@href}%
\providecommand \@href[1]{\@@startlink{#1}\@@href}%
\providecommand \@@href[1]{\endgroup#1\@@endlink}%
\providecommand \@sanitize@url [0]{\catcode `\\12\catcode `\$12\catcode `\&12\catcode `\#12\catcode `\^12\catcode `\_12\catcode `\%12\relax}%
\providecommand \@@startlink[1]{}%
\providecommand \@@endlink[0]{}%
\providecommand \url  [0]{\begingroup\@sanitize@url \@url }%
\providecommand \@url [1]{\endgroup\@href {#1}{\urlprefix }}%
\providecommand \urlprefix  [0]{URL }%
\providecommand \Eprint [0]{\href }%
\providecommand \doibase [0]{https://doi.org/}%
\providecommand \selectlanguage [0]{\@gobble}%
\providecommand \bibinfo  [0]{\@secondoftwo}%
\providecommand \bibfield  [0]{\@secondoftwo}%
\providecommand \translation [1]{[#1]}%
\providecommand \BibitemOpen [0]{}%
\providecommand \bibitemStop [0]{}%
\providecommand \bibitemNoStop [0]{.\EOS\space}%
\providecommand \EOS [0]{\spacefactor3000\relax}%
\providecommand \BibitemShut  [1]{\csname bibitem#1\endcsname}%
\let\auto@bib@innerbib\@empty
\bibitem [{\citenamefont {Parakkat}\ \emph {et~al.}(2016)\citenamefont {Parakkat}, \citenamefont {Ganesh},\ and\ \citenamefont {Anil~Kumar}}]{parakkat2016tailoring}%
  \BibitemOpen
  \bibfield  {author} {\bibinfo {author} {\bibfnamefont {V.~M.}\ \bibnamefont {Parakkat}}, \bibinfo {author} {\bibfnamefont {K.}~\bibnamefont {Ganesh}},\ and\ \bibinfo {author} {\bibfnamefont {P.}~\bibnamefont {Anil~Kumar}},\ }\bibfield  {title} {\bibinfo {title} {Tailoring curie temperature and magnetic anisotropy in ultrathin $\mathrm{Pt}/\mathrm{Co}/\mathrm{Pt}$ films},\ }\href {https://doi.org/10.1063/1.4944343} {\bibfield  {journal} {\bibinfo  {journal} {AIP Advances}\ }\textbf {\bibinfo {volume} {6}},\ \bibinfo {pages} {056118} (\bibinfo {year} {2016})}\BibitemShut {NoStop}%
\bibitem [{\citenamefont {Kanak}\ \emph {et~al.}(2007)\citenamefont {Kanak}, \citenamefont {Czapkiewicz}, \citenamefont {Stobiecki}, \citenamefont {Kachel}, \citenamefont {Sveklo}, \citenamefont {Maziewski},\ and\ \citenamefont {Van~Dijken}}]{kanak2007influence}%
  \BibitemOpen
  \bibfield  {author} {\bibinfo {author} {\bibfnamefont {J.}~\bibnamefont {Kanak}}, \bibinfo {author} {\bibfnamefont {M.}~\bibnamefont {Czapkiewicz}}, \bibinfo {author} {\bibfnamefont {T.}~\bibnamefont {Stobiecki}}, \bibinfo {author} {\bibfnamefont {M.}~\bibnamefont {Kachel}}, \bibinfo {author} {\bibfnamefont {I.}~\bibnamefont {Sveklo}}, \bibinfo {author} {\bibfnamefont {A.}~\bibnamefont {Maziewski}},\ and\ \bibinfo {author} {\bibfnamefont {S.}~\bibnamefont {Van~Dijken}},\ }\bibfield  {title} {\bibinfo {title} {Influence of buffer layers on the texture and magnetic properties of $\mathrm{Co}/\mathrm{Pt}$ multilayers with perpendicular anisotropy},\ }\href {https://doi.org/10.1002/pssa.200777104} {\bibfield  {journal} {\bibinfo  {journal} {Physica Status Solidi (a)}\ }\textbf {\bibinfo {volume} {204}},\ \bibinfo {pages} {3950} (\bibinfo {year} {2007})}\BibitemShut {NoStop}%
\bibitem [{\citenamefont {Yıldırım}\ \emph {et~al.}(2022)\citenamefont {Yıldırım}, \citenamefont {Marioni}, \citenamefont {Falub}, \citenamefont {Rohrmann}, \citenamefont {Jaeger}, \citenamefont {Rechsteiner}, \citenamefont {Schneider},\ and\ \citenamefont {Hug}}]{yildirim2022tuning}%
  \BibitemOpen
  \bibfield  {author} {\bibinfo {author} {\bibfnamefont {O.}~\bibnamefont {Yıldırım}}, \bibinfo {author} {\bibfnamefont {M.~A.}\ \bibnamefont {Marioni}}, \bibinfo {author} {\bibfnamefont {C.~V.}\ \bibnamefont {Falub}}, \bibinfo {author} {\bibfnamefont {H.}~\bibnamefont {Rohrmann}}, \bibinfo {author} {\bibfnamefont {D.}~\bibnamefont {Jaeger}}, \bibinfo {author} {\bibfnamefont {M.}~\bibnamefont {Rechsteiner}}, \bibinfo {author} {\bibfnamefont {D.}~\bibnamefont {Schneider}},\ and\ \bibinfo {author} {\bibfnamefont {H.~J.}\ \bibnamefont {Hug}},\ }\bibfield  {title} {\bibinfo {title} {Tuning the perpendicular magnetic anisotropy in $\mathrm{Co}/\mathrm{Pt}$ multilayers grown by facing target sputtering and conventional sputtering},\ }\href {https://doi.org/10.1016/j.scriptamat.2021.114285} {\bibfield  {journal} {\bibinfo  {journal} {Scripta Materialia}\ }\textbf {\bibinfo {volume} {207}},\ \bibinfo {pages} {114285} (\bibinfo {year} {2022})}\BibitemShut {NoStop}%
\bibitem [{\citenamefont {Carcia}(1988)}]{Carcia1988}%
  \BibitemOpen
  \bibfield  {author} {\bibinfo {author} {\bibfnamefont {P.~F.}\ \bibnamefont {Carcia}},\ }\bibfield  {title} {\bibinfo {title} {Perpendicular magnetic anisotropy in $\mathrm{Pd}/\mathrm{Co}$ and $\mathrm{Pt}/\mathrm{Co}$ thin-film layered structures},\ }\href {https://doi.org/10.1063/1.340404} {\bibfield  {journal} {\bibinfo  {journal} {Journal of Applied Physics}\ }\textbf {\bibinfo {volume} {63}},\ \bibinfo {pages} {5066} (\bibinfo {year} {1988})}\BibitemShut {NoStop}%
\bibitem [{\citenamefont {Mangin}\ \emph {et~al.}(2006)\citenamefont {Mangin}, \citenamefont {Ravelosona}, \citenamefont {Katine}, \citenamefont {Carey}, \citenamefont {Terris},\ and\ \citenamefont {Fullerton}}]{mangin2006current}%
  \BibitemOpen
  \bibfield  {author} {\bibinfo {author} {\bibfnamefont {S.}~\bibnamefont {Mangin}}, \bibinfo {author} {\bibfnamefont {D.}~\bibnamefont {Ravelosona}}, \bibinfo {author} {\bibfnamefont {J.}~\bibnamefont {Katine}}, \bibinfo {author} {\bibfnamefont {M.}~\bibnamefont {Carey}}, \bibinfo {author} {\bibfnamefont {B.}~\bibnamefont {Terris}},\ and\ \bibinfo {author} {\bibfnamefont {E.~E.}\ \bibnamefont {Fullerton}},\ }\bibfield  {title} {\bibinfo {title} {Current-induced magnetization reversal in nanopillars with perpendicular anisotropy},\ }\href {https://doi.org/10.1038/nmat1595} {\bibfield  {journal} {\bibinfo  {journal} {Nature Materials}\ }\textbf {\bibinfo {volume} {5}},\ \bibinfo {pages} {210} (\bibinfo {year} {2006})}\BibitemShut {NoStop}%
\end{thebibliography}%

\end{document}


\preprint{Ojha {\it et al.}, Tailoring the Topological Hall Effect in Pt/Co/X (X = Ta, Re) thin films}

\title{ Supplementary Information\\
\vspace{0.5cm}
Tailoring the Topological Hall Effect in Pt/Co/X (X = Ta, Re) thin films}

\author {Brindaban Ojha}
\author {Shaktiranjan Mohanty} 
\author {Bhuvneshwari Sharma}
\author{Subhankar Bedanta}
\email{sbedanta@niser.ac.in}
\address {Laboratory for Nanomagnetism and Magnetic Materials (LNMM), School of Physical Sciences, National Institute of Science Education and Research (NISER),  An OCC of Homi Bhabha National Institute (HBNI), Jatni 752050, Odisha, India}


\pacs{}

\maketitle

\onecolumngrid

\section{Structural characterization}

\begin{figure} [H]
	\centering
	\includegraphics[width=0.5\linewidth]{"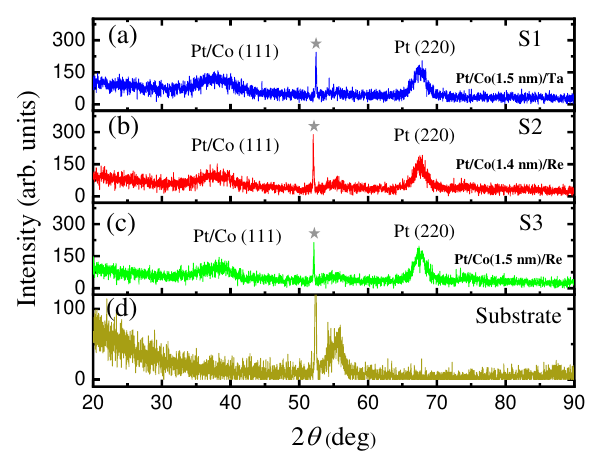"}
    \renewcommand{\thefigure}{S\arabic{figure}}
	\caption{(a) - (d) GIXRD measurements of samples S1, S2, S3 and substrate, respectively. The stars in the figure represent the peak that arises from the substrate.  }
	\label{fig:Fig_S1}
\end{figure}

In the Grazing incidence X-ray diffraction (GIXRD) measurements, we have observed the (111) and (220) planes of Pt at 2$\theta$ $\approx$ 38$^{\circ}$ and 67.7$^{\circ}$ \cite{parakkat2016tailoring}, as shown in Fig. \ref{fig:Fig_S1} (a-c). However, the (111) peaks are very broad and can be considered as fcc Pt/Co (111) reflection \cite{kanak2007influence,yildirim2022tuning}. Thus, it could be concluded that the films (S1, S2, and S3) are textured with a (111) orientation along the growth direction \cite{Carcia1988,kanak2007influence,mangin2006current}. 

\begin{figure} [H]
	\centering
	\includegraphics[width=1.0\linewidth]{"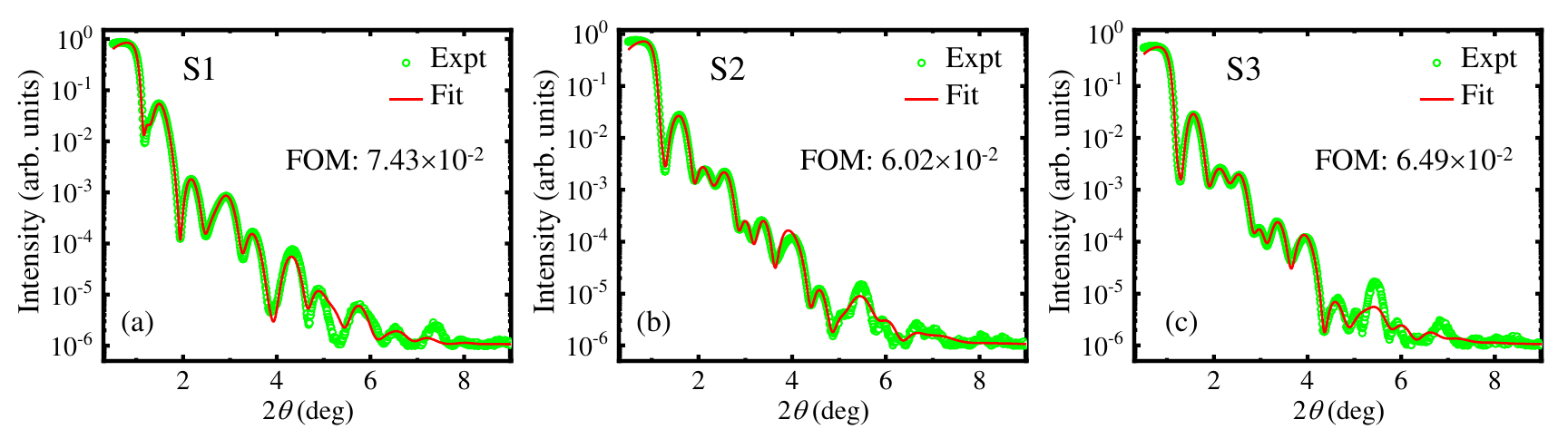"}
    \renewcommand{\thefigure}{S\arabic{figure}}
	\caption{ XRR fits for samples S1, S2, and S3 are shown in (a), (b), and (c), respectively.  }
	\label{fig:Fig_S2}
\end{figure}

The Fig. \ref{fig:Fig_S2} (a), (b) and (c) shows data and fit of X-ray reflectivity (XRR) measurements of sample S1, S2 and S3, respectively. The green circles and solid red lines indicate the measured experimental data and corresponding fittings. The thickness and roughness are tabulated in the Table \ref{table:tableS_1}. The Co thickness is found to be $\sim$ 1.5, 1.4 and 1.5 nm for sample S1, S2 and S3, respectively.

\begin{table*} [t]
\renewcommand{\thetable}{S\arabic{table}}
\caption{Thickness and roughness of Pt/Co/Ta and Pt/Co/Re samples as obtained from XRR fits. The figure of merit (FOM) of the fits are $7.43 \times 10^{-2}$, $6.02 \times 10^{-2}$, and $6.49 \times 10^{-2}$ for S1-S3, respectively} 
	\label{tbl:example}
\begin{tabular}{|l|l|l|l|}
\hline Layers & Sample S1 (in nm) & Sample S2 (in nm) & Sample S3 (in nm)\\
\hline Ta (bottom) thickness & 4.84 & 5.64 & 5.72 \\
\hline Ta (bottom) roughness & 0.76 & 0.50 & 0.56 \\
\hline Pt thickness & 5.46 & 5.76 & 5.60 \\
\hline Pt roughness & 1.08 & 0.80 & 0.76 \\
\hline Co thickness & 1.49 & 1.40 & 1.51 \\
\hline Co roughness & 0.41 & 0.59 & 0.56 \\
\hline Re thickness & - & 3.51 & 3.45 \\
\hline Re roughness & - & 0.68 & 0.67 \\
\hline Ta (top) thickness & 4.18 & 3.06 & 2.92 \\
\hline Ta (top) roughness & 1.29 & 0.83 & 0.82 \\
\hline TaO$_x$ (top) thickness & 2.33 & 2.64 & 3.33 \\
\hline TaO$_x$ (top) roughness & 0.54 & 0.79 & 0.95 \\
\hline
\end{tabular}
\label{table:tableS_1}
\end{table*}

\section{Magnetometry}

\begin{figure} [H]

\centering\includegraphics[width=0.4\linewidth]{"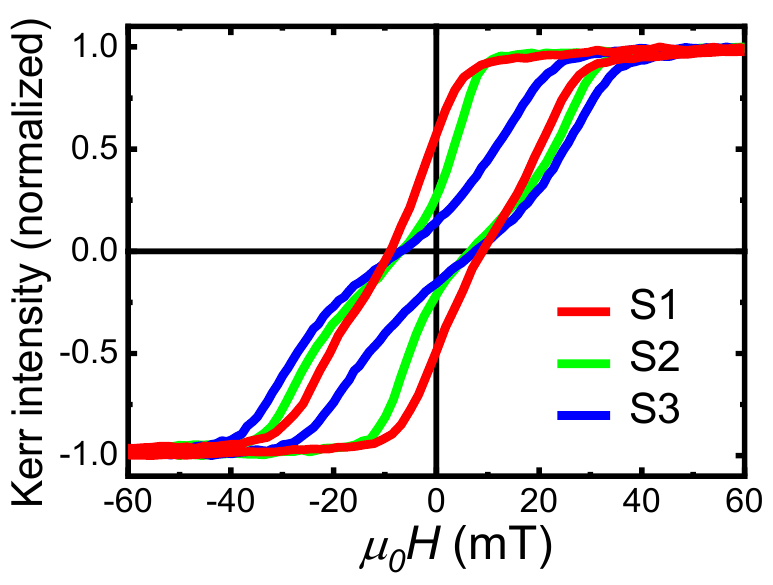"}
    \renewcommand{\thefigure}{S\arabic{figure}}
	\caption{ Polar MOKE hysteresis loops of the samples S1-S3.
    }
	\label{fig:Fig_S3}
\end{figure}

Fig. \ref{fig:Fig_S3} shows the polar MOKE hysteresis loops, confirming that the easy axis of all the samples aligns in the OOP direction.

\begin{figure} [H]
	\centering
	\includegraphics[width=1.0\linewidth]{"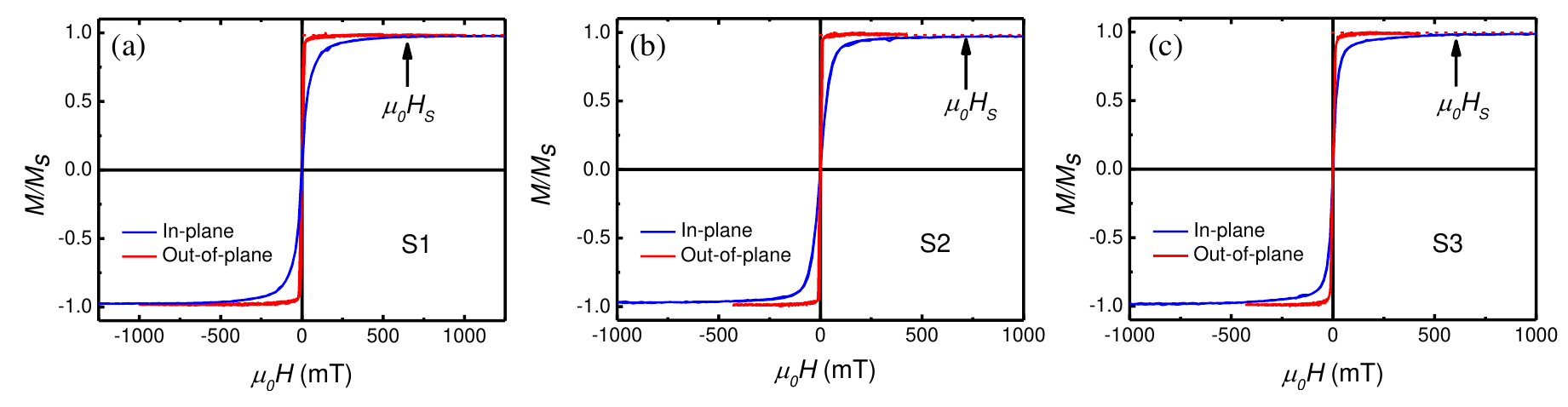"}
    \renewcommand{\thefigure}{S\arabic{figure}}
	\caption{ The out-of-plane and in-plane hysteresis loops of S1, S2 and S3 are shown in (a), (b) and (c), respectively.}
	\label{fig:Fig_S4}
\end{figure}

Fig. \ref{fig:Fig_S4} (a-c) reveals the out-of-plane and inplane SQUID loops of S1, S2 and S3, respectively. The arrow signs indicate the in-plane saturation field ($H_s$) of the samples.

\begin{figure} [H]
	\centering
	\includegraphics[width=0.8\linewidth]{"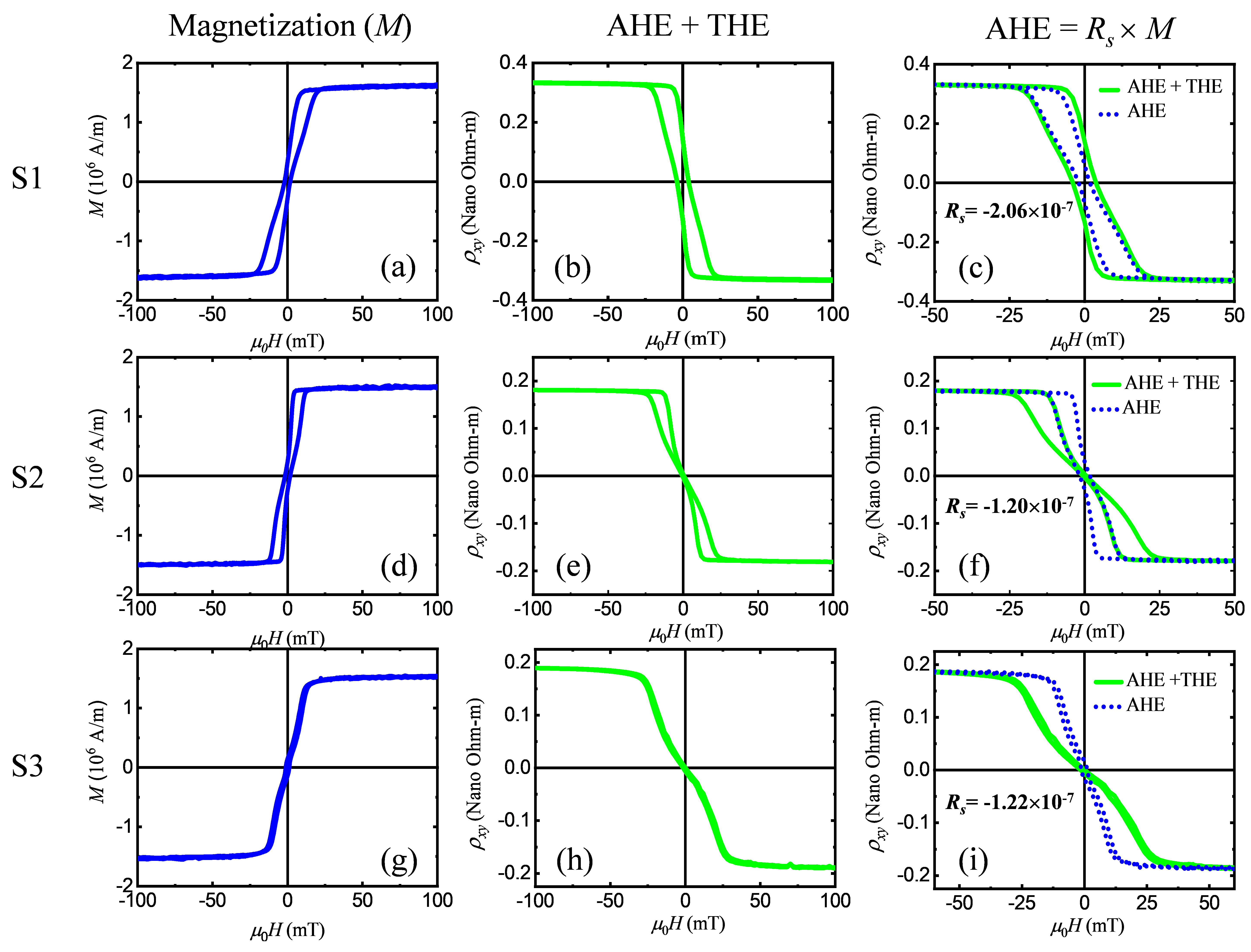"}
    \renewcommand{\thefigure}{S\arabic{figure}}
	\caption{ (a), (d), and (g) M-H loops, and (b), (e), and (h) Hall resistivity as a function of field for samples S1, S2, and S3, respectively. To quantify the AHE contribution, the magnetization is scaled by $R_s$. The scaled $M_s$ (or AHE) is plotted with Hall resistivity (AHE + THE) in (c), (f), and (i) for S1–S3, respectively. The unit of $R_s$ is nano Ohm-m$^2$/A. }
	\label{fig:Fig_S5}
\end{figure}

Fig. \ref{fig:Fig_S5} (a), (d) and (g) represent M-H loops of the samples S1, S2, and S3, respectively. Figures S3 (b), (e) and (h) reveal Hall resistivity (AHE + THE) loops of the samples S1, S2, and S3, respectively. Now, to extract the AHE, the M-H loops are scaled by multiplying the $R_s$ =${\rho_{xy, saturation}}/{M_s}$, here $R_s$ and $M_s$ are the anomalous Hall coefficient and saturation magnetization, respectively, as shown in figure S3 (c), (f) and (i) of the samples S1, S2, and S3, respectively.

\section{MuMax simulations}

\begin{figure} [H]
	\centering
	\includegraphics[width=0.7\linewidth]{"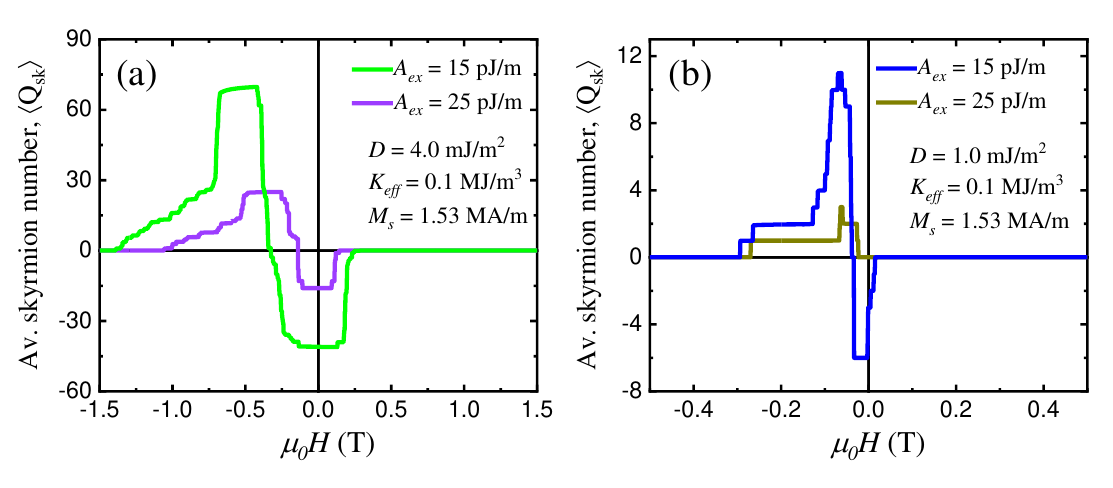"}
    \renewcommand{\thefigure}{S\arabic{figure}}
	\caption{ A double peak with opposite sign in $\left\langle Q_{sk} \right\rangle$ for varying $A_{ex}$ with (a) $D$ = 4.0 mJ/m$^2$, $K_{eff}$ = 0.1 MJ/m$^3$, and $M_s$ =1.53 MA/m, and (b) $D$ = 1.0 mJ/m$^2$, $K_{eff}$ = 0.1 MJ/m$^3$, and $M_s$ =1.53 MA/m, obtained from simulations.}
	\label{fig:Fig_S6}
\end{figure}

Upon decreasing $A_{ex}$ to 15 pJ/m, the double peaks always persist in $\left\langle Q_{sk} \right\rangle$ with reducing $K_{eff}$ (Fig. \ref{fig:Fig_S6}(a)) or simultaneously lowering both $D$ and $K_{eff}$ (see Fig. \ref{fig:Fig_S6}(b)).

\begin{figure} [H]
	\centering
	\includegraphics[width=1.0\linewidth]{"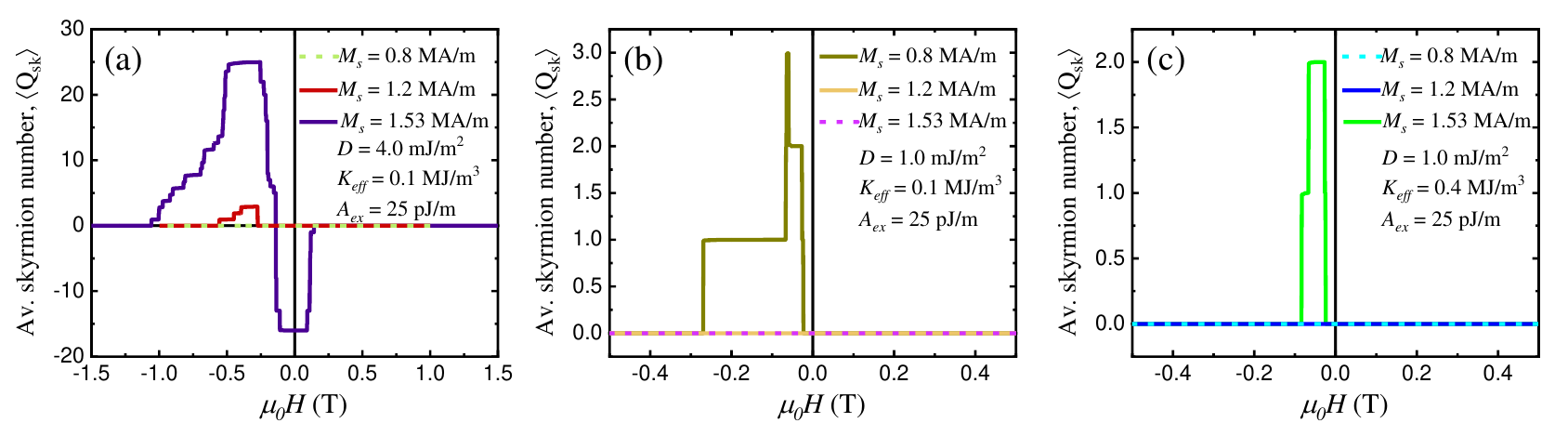"}
    \renewcommand{\thefigure}{S\arabic{figure}}
	\caption{ Evolution from a double peak to no peak in $\left\langle Q_{sk} \right\rangle$ for (a) $D$ = 4.0 mJ/m$^2$, $K_{eff}$ = 0.1 MJ/m$^3$, $A_{ex}$ = 25 pJ/m and from a single peak to no peak for (b) $D$ = 1.0 mJ/m$^2$, $K_{eff}$ = 0.1 MJ/m$^3$, $A_{ex}$ = 25 pJ/m, (c) $D$ = 1.0 mJ/m$^2$, $K_{eff}$ = 0.4 MJ/m$^3$, $A_{ex}$ = 25 pJ/m with varying $M_s$.}
	\label{fig:Fig_S7}
\end{figure}

With reducing the $K_{eff}$ to 0.1 MJ/m$^3$, $\left\langle Q_{sk} \right\rangle$ exhibits a transition from a double peak to a single peak and eventually to no peak as $M_s$ decreases (see Fig. \ref{fig:Fig_S7}(a)). Additionally, we examined the effect of reducing $D$ to 1 mJ/m$^2$ while considering $K_{eff}$  values of 0.1 MJ/m$^3$ and 0.4 MJ/m$^3$. For each combination of D and $K_{eff}$, we systematically varied $M_s$ (see Fig. \ref{fig:Fig_S7}(b-c)). In both cases, a single peak is observed for $M_s$ = 1.53 MA/m, while further reduction in $M_s$ results in the suppression of the peak (see Fig. \ref{fig:Fig_S7}(b-c)).

\bibliography{References}